\documentclass[lettersize,journal]{IEEEtran}

\usepackage{amsmath}
\usepackage{pifont}

\usepackage{comment}
\usepackage{algpseudocode}
\usepackage{algorithmicx,algorithm}
\usepackage{url} 

\usepackage{wrapfig}
\usepackage{diagbox}
\usepackage{enumitem}
\usepackage{multirow}
\usepackage{authblk}
\usepackage{fancyhdr}
\usepackage[numbers]{natbib}

\usepackage{hyperref} 
\usepackage{graphicx}
\usepackage{amssymb}
\usepackage{cite}
\begin{document}
\title{A Learning-Based Attack Framework to Break SOTA Poisoning Defenses in Federated Learning}

\author{Yuxin~Yang,
        Qiang~Li,
        Chenfei~Nie,
        Yuan~Hong,
        Meng~Pang,
        and  Binghui~Wang
\thanks{Yuxin Yang is with the College of Computer Science and Technology, Jilin University, China, and Department of Computer Science, Illinois Institute of Technology, USA. Qiang Li and Chenfei Nie are with the College of Computer Science and Technology, Jilin University, China. Yuan Hong is with the School of Computing at University of Connecticut, USA. Meng Pang is with the School of Mathematics and Computer Sciences, Nanchang University, China. Binghui Wang is with the Department of Computer Science, Illinois Institute of Technology, USA. Wang is the corresponding author (bwang70@iit.edu).}
}

% The paper headers
\markboth{ }%
{Shell \MakeLowercase{\textit{et al.}}: Bare Demo of IEEEtran.cls for IEEE Journals}

% 
\iffalse
\author{
First Author$^1$
\and
Second Author$^2$\and
Third Author$^{2,3}$\And
Fourth Author$^4$
\affiliations
$^1$First Affiliation\\
$^2$Second Affiliation\\
$^3$Third Affiliation\\
$^4$Fourth Affiliation
\emails
\{first, second\}@example.com,
third@other.example.com,
fourth@example.com
}
\fi

\maketitle
\begin{abstract}
Federated Learning (FL) is a novel client-server distributed learning framework  
that can 
protect data privacy. 
However, recent works show that FL is vulnerable to poisoning attacks. Many defenses with robust aggregators (AGRs) are proposed  to mitigate the issue, but they are all broken by advanced attacks. 
Very recently, some renewed robust AGRs are designed, typically with novel clipping or/and filtering strategies, and they show promising defense performance against the advanced poisoning attacks. 
In this paper, we show that these novel robust AGRs are also vulnerable to  carefully designed poisoning attacks. 
Specifically, we observe that breaking these robust AGRs reduces to bypassing the clipping or/and filtering of malicious clients, and propose an optimization-based attack framework to leverage this observation. 
Under the framework, we then design the customized attack 
against each robust AGR. 
Extensive experiments on multiple datasets and threat models verify 
our proposed optimization-based attack can break the  SOTA AGRs. 
We hence call for novel defenses against poisoning attacks to FL. Code is available at: \url{https://github.com/Yuxin104/BreakSTOAPoisoningDefenses}. 

\end{abstract}


\section{Introduction}

\IEEEPARstart{F}{ederated} Learning (FL) \citep{mcmahan2017communication, sheller2019multi,li2020lotteryfl,yang2024distributed}, a novel client-server distributed learning {paradigm}, allows 
participating clients keep and train their data locally 
and only share the trained local models (e.g., gradients), instead of the raw data, 
with a center server for aggregation. 
The aggregated local models forms a shared global model, which is used by clients for their main task. FL thus has been a great potential to protect data privacy and 
is widely applied to medical \citep{sohan2023systematic}, financial \citep{li2022nearest}, and other privacy-sensitive applications such as on-client item ranking~\citep{mcmahan2017communication}, content suggestions for on-device keyboards~\citep{bonawitz2019towards}, and next word prediction~\citep{li2020federated}. However, recent works show that the invisibility of client data also renders FL vulnerable to \emph{poisoning attacks}~\citep{baruch2019little,bhagoji2019analyzing,fang2020local,shejwalkar2021manipulating,saha2020hidden,bagdasaryan2020backdoor,wang2020attack,xie2020dba,zhang2022neurotoxin,10549523}, which aim to  manipulate the training phase (and testing phase) of FL to disrupt the global model behavior or/and degrade the FL performance. 

To defend against the 
poisoning attacks to FL, numerous robust aggregation algorithms (AGRs) 
\citep{blanchard2017machine,chen2017distributed,guerraoui2018hidden,yin2018byzantine,xie2018generalized,chen2018draco,li2019rsa,pillutla2019robust,xie2019zeno,munoz2019byzantine,wu2020federated,cao2020fltrust,rieger2022deepsight,nguyen2021flguard} 
have been proposed, where the key idea is to design a robust aggregator that aims to filter malicious gradients, i.e., 
those largely deviate from others. 
For instance,  the Krum AGR \citep{blanchard2017machine} selects the gradient that is closest to its $n-f-2$ neighboring gradients measured by Euclidean distance, in order to filter $f$ malicious gradients and $n$ is the total number of clients. 
However, these AGRs  are all 
broken by an advanced 
attack \citep{shejwalkar2021manipulating}. 
To 
further address the issue, 
some renewed 
AGRs equipped with an enhanced defense ability are proposed, 
which include 
FLAME \citep{nguyenflame}, 
MDAM \citep{farhadkhani2022byzantine}, {FLDetector \citep{zhang2022fldetector}} and Centered Clipping (CC) \citep{karimireddy2021learning} (or its variant Bucketing \citep{karimireddy2020byzantine} to handle non-IID data). 
These new robust AGRs have shown effective defense performance against the advanced poisoning attack.  

\begin{figure*}[!t]
\centering	
\includegraphics[width=0.98\textwidth]{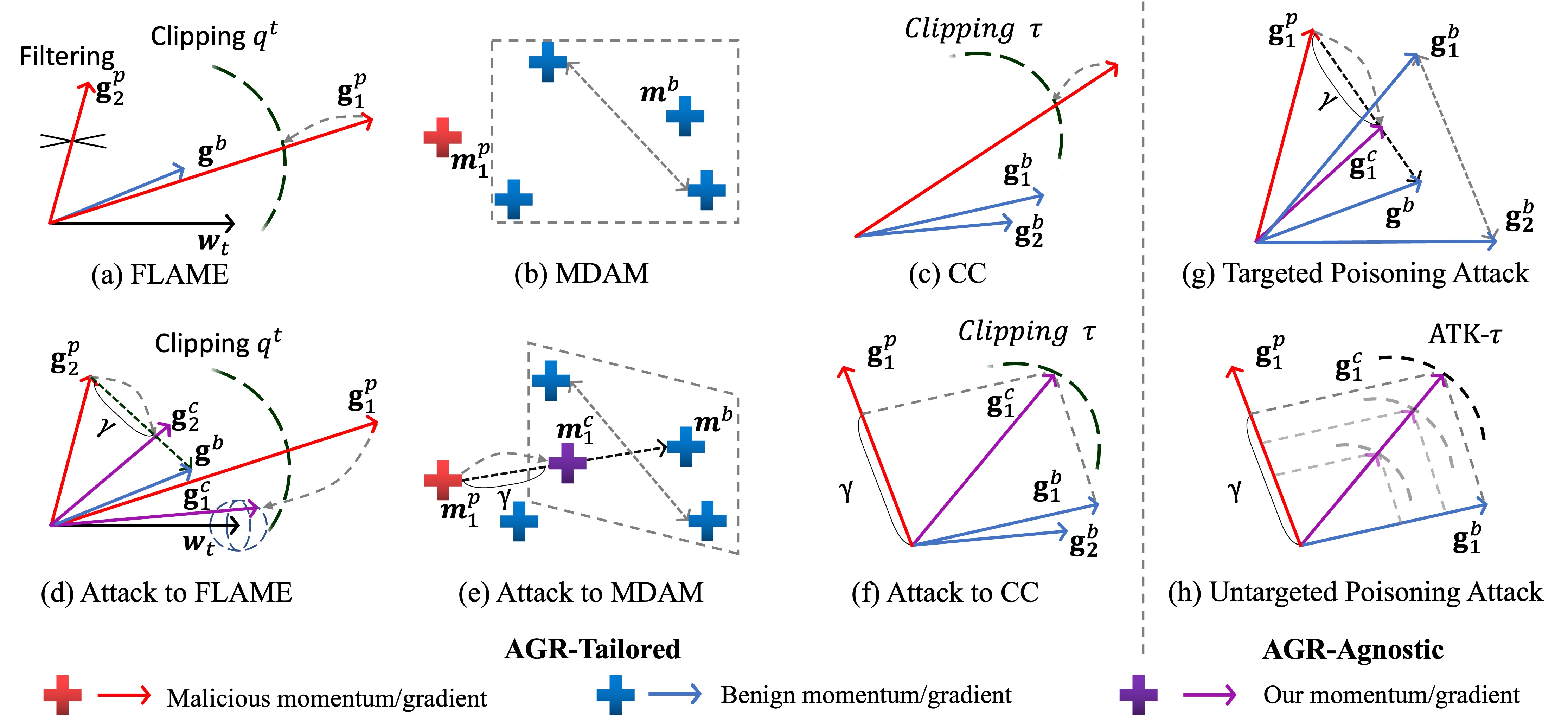}
	\caption{Illustration of the SOTA robust aggregation algorithms (a)-(c) in FL, our AGR-tailored attacks (d)-(f) {and AGR-agnostic attacks (g)-(h)} on them. (a) {\bf FLAME:} it defends against the malicious gradients via  clipping and filtering gradients that with high length and angular deviations, respectively. (b) {\bf MDAM:} it chooses a subset of $ n-f $ momentums with the smallest diameter for aggregation, i.e., filter out a bounded number of $f$ malicious gradients.   
 (c) {\bf CC:} it  corrects malicious gradients via a centered clipping with a parameter $\tau$. (d) {\bf Our attack to FLAME:} we project the length-deviating malicious gradients and rotate the angle-deviating malicious gradients to evade FLAME. (e) {\bf Our attack to MDAM:} we optimize the original malicious momentums to new ones such that MDAM selects (part of) the new malicious ones into the subset for aggregation. (f) {\bf Our attack to CC:} we construct malicious gradients from any benign one to avoid the center clipping. {(g) Our AGR-agnostic targeted poisoning attack (on FLAME, MDAM, and FLDetector): we adjust malicious gradients to approach benign gradients, based on the Euclidean distance metric, to evade SOTA defenses. (h) Our AGR-agnostic untargeted poisoning attack (to CC): we generate malicious gradients of length ATK-$\tau$ by leveraging any of benign gradients to evade clipping for agnostic parameters.
 }} 
        \label{fig:overview}
\end{figure*}

In this paper, however, we demonstrate that all these 
SOTA AGRs are still vulnerable to \emph{carefully designed} (untargeted and targeted) poisoning attacks. 
Specifically, we first scrutinize these AGRs and find that they adopt either a novel \emph{clipping} (e.g., in CC) or \emph{filtering} (e.g., in MDAM {and FLDetector}), or the both (e.g., in FLAME) to remove the effect caused by malicious clients. Breaking these robust AGRs hence reduces to making malicious clients bypassing the clipping or/and filtering (See Figure \ref{fig:overview}). 
We then formalize this observation 
and design an optimization-based attack framework. 
To be specific,  
we first 
analyze the inherent attack objective and constraint within different AGRs and threat models. 
Then we instantiate this framework by: 1) adjusting the initial malicious gradients for targeted poisoning attacks to adhere to the constraint, and 2) constructing malicious gradients from benign ones for untargeted poisoning attacks to satisfy this constraint, thereby evading the filtering and clipping mechanisms. 
Finally, we propose an efficient solution for the optimization problem\footnote{We emphasize that 
our attack framework differs with \citep{shejwalkar2021manipulating} and extends 
it in two key aspects. First, the AGRs under consideration are more recent and robust than those considered in 
\citep{shejwalkar2021manipulating}. That is, all AGRs successfully  broken by \citep{shejwalkar2021manipulating} can be also broken by our framework, while the opposite is not true. 
Second, our framework encompasses both targeted and untargeted attacks on robust AGRs, 
while \citep{shejwalkar2021manipulating} only focuses on untargeted attacks. Note that the generalization on targeted attacks is non-trivial. For instance, {we test that the untargeted poisoning attack based on \citep{shejwalkar2021manipulating} completely fails to break 
FLAME, MDAM, {and FLDetector}, while our designed targeted poisoning attacks can be successful (see {Table \ref{FLAME-results}}, Table \ref{MDAM-results}, and {Table \ref{FLDetector-results}}).}}.  

We extensively investigate the effectiveness of our attack framework  against the SOTA robust AGRs 
under multiple experimental settings and datasets. 
Our empirical evaluations show that our proposed poisoning attacks vigorously disrupt these robust AGRs.  
Below, we summarize the main contributions as follows:
\begin{itemize}[leftmargin=*]
\item We show that state-of-the-art AGRs can still be broken by advanced attacks. 
\item 
{We propose an optimization-based attack framework that explores both targeted and untargeted poisoning attacks against SOTA defenses under diverse scenarios.}

\item We validate the effectiveness of our attacks on multiple experimental settings and datasets. 
\end{itemize}

\section{Preliminaries}
\label{sec:prelim}
\subsection{Federated Learning (FL)} 
FL 
links a set of (e.g., $n$) clients and a server to collaboratively and iteratively train a shared global model over private client data, 
where the data across clients may be non independently and identically distributed (non-IID). Specifically, in a $t$-th round, 
 the server selects a subset of clients $S^t \subset [n]$ and broadcasts the current global model parameters, denoted as ${\bf w}^t$, to the chosen clients $S^t$. 
Each chosen client $i \in S^t$ then calculates the gradient 
$\mathbf{g}_i^{t} = \partial_{{\bf w}^t} L(D_i; {\bf w}^t)$ using its local data $D_i$ and sends $\mathbf{g}_i^{t}$ to the server. Here $L$ is the loss function, e.g., cross-entropy loss. 
The server aggregates the collected clients' gradients using some aggregation algorithm $\textrm{AGR}(\mathbf{g}_{i \in [n]}^{t})$, e.g., dimension-wise average in the well-known FedAvg~\citep{mcmahan2017communication} where $\textrm{AGR}(\mathbf{g}_{i \in [n]}^{t}) = \frac{1}{|S^t|}
\sum_{i \in S^t}\mathbf{g}_i^t$.  
Finally, the server updates the global model for the next round ${\bf w}^{t+1}$ using the aggregator and SGD, e.g.,  
${\bf w}^{t+1}={\bf w}^t - \eta \textrm{AGR}(\mathbf{g}_{i \in [n]}^{t})$ with a learning rate $\eta$, and broadcasts it to a new subset of randomly chosen clients $S^{t+1}$. 
This process is repeated until the global model converges or reaches the maximal round. 
The final global model is used by clients for their main task. 

\subsection{SOTA Robust Aggregators for FL}
\label{sec:SOTAAGEs}
In this section, we will review the {four} SOTA robust AGRs 
for FL against poisoning attacks. 
{The detailed implementations of these AGRs are shown in Algorithms~\ref{FLAME-algorithm}-\ref{CC-B-algorithm}} in Appendix~\ref{app:code}.

\vspace{+0.05in}
\noindent {\bf FLAME~\citep{nguyenflame}.} It is a defense against targeted poisoning attacks, particularly backdoor attacks. 
FLAME is based on the intuition that  malicious gradients tend to deviate from the benign ones in length and/or angle. 
To limit the impact of backdoored models,  
FLAME proposes a two-stage solution in each  $t$-th round. 
First, it clips the client gradient with length larger than a clipping threshold $q^t$, i.e., $\mathbf{g}_i^t \gets \frac{q^t}{\|\mathbf{g}_i^t\|_2} \mathbf{g}_i^t$, if $\|\mathbf{g}_i^t\|_2 > q^t, \forall i$. 
FLAME chooses $q^t$ as $q^t = \textrm{MEDIAN}(\|\mathbf{g}_1\|_2^t,...,\|\mathbf{g}_n\|_2^t)$. 
Second, 
it inputs the clipped gradients and utilizes a dynamic clustering technique HDBSCAN \citep{campello2013density} to further filter out gradients with high angular deviations from the majority gradients. Specifically, 
HDBSCAN  clusters the clients based on the density of cosine distance distribution on gradient pairs and dynamically determines the  number of clusters. By assuming an upper bounded ($<n/2$) number of malicious clients, FLAME sets the minimum size of the largest cluster to be ${n}/{2}+1 $ to ensure that the resulting cluster only contains benign clients. 
With the  
clipping and filtering, FLAME finally only  
averages the gradients of clients in the benign cluster to update the global model. 

\vspace{+0.05in}
\noindent {\bf MDAM~\citep{farhadkhani2022byzantine}.} 
It proposes to 
use distributed momentum into existing aggregators (e.g., Krum, Median, TM, MDA) to enhance the robustness against poisoning attacks, and provably shows that MDA under distributed momentum (MDAM) obtains the best defense performance. 
Specifically, at a $t$-th round, each client $i$ sends to the server the momentum $\mathbf{m}^t_i=\beta \mathbf{m}^{t-1}_i+(1-\beta)\mathbf{g}^t_i$, instead of the gradient $\mathbf{g}^t_i$, where the initial momentum is $\mathbf{m}^0_i=0$ and $\beta \in [0,1)$ is the momentum coefficient. After the server receives %the 
$n$ momentums $\{{\bf m}^t_i\}$, it first decides a set $S^t$ of $ n-f $ clients with the smallest diameter, i.e., $S^t \in \underset{ S \subset [n], |S|=n-f }{\arg\min} \left\{ \underset{i,j \in S}{\max}\left \|\mathbf{m}^t_i-\mathbf{m}^t_j  \right \|_2 \right\}$, 
where at most $f$ clients are assumed  malicious; and then updates the servers' global model as ${\bf w}^{t+1}={\bf w}^t+\frac{1}{n-f}\sum_{i \in S^t}\mathbf{m}^t_i$. 

\vspace{+0.05in}
\noindent {{\bf FLDetector \citep{zhang2022fldetector}.}  
It aims at detecting and removing 
malicious clients by assessing the (in)consistency of clients' model updates in each round. 
Specifically, in a $t$-th round, the server predicts a client $i$'s model update $\hat{g}_i^t$ using the historical global model updates and the estimated Hessian $\hat{H}^t$, i.e., $\hat{g}_i^t = g_i^{t-1}+\hat{H}^t({\bf w}_t-{\bf w}_{t-1})$. The suspicious score $s_i^t$ for client $i$ is the client’s average normalized Euclidean distance in the past $N$ iterations, i.e., $s_i^t = \frac{1}{N} {\textstyle \sum_{r=0}^{N-1}} d_i^{t-r}/\left \|d^{t-r} \right \|_1$, where $d^t = [\left \|\hat{g}_1^t - g_1^t \right \|_2,\left \|\hat{g}_2^t - g_2^t  \right \|_2,...,\left \|\hat{g}_n^t - g_n^t  \right \|_2]$. Finally, FLDetector utilizes $k$-means with Gap statistics \citep{tibshirani2001estimating} on the clients’ suspicious scores to detect and filter malicious clients.}

\vspace{+0.05in}
\noindent {\bf Centered Clipping (CC)~\citep{karimireddy2021learning}.} It is a simple and efficient clipping based robust AGR which provides a standardized specification for ``robust'' robust aggregators. To circumvent poisoning attacks, CC clips each 
(benign or malicious) client gradient $\mathbf{g}_i^t$ in each round $t$ by $\mathbf{g}_i^{t}=\mathbf{g}_i^{t} \cdot \min (1,\frac{\tau}{\left \| \mathbf{g}_i^{t} \right \|_2 } )$, 
where $\tau$ is a predefined clipping threshold. 
Then the global model 
is updated as $ {\bf w}^{t+1}={\bf w}^t+{\eta}/{n}\sum_{i=1}^{n}\mathbf{g}_i^t$. Unlike the majority of other AGRs, 
CC is very scalable and requires only $O(n)$ computation and communication cost per round.

\noindent {\bf Centered Clipping with Bucketing (CC-B).}  To further adapt the robust CC AGR to heterogeneous (non-IID) datasets, \citep{karimireddy2020byzantine} proposes a bucketing scheme to ``mix'' the data from all clients, which
reduces the chance of any subset of the client data being consistently ignored. To be specific, in a $t$-th round, it first generates a random permutation $\pi$ of $[n]$, and computes $\bar{\bf g}^t_i=\frac{1}{s} \sum_{k=(i-1)\cdot s+1}^{\min(n,i \cdot s)} \mathbf{g}^t_{\pi (k)}$ for $i=\{ 1,...,\left \lceil  n/s\right \rceil \}$, where $s$ is the number of buckets. Then it combines with the CC AGR to update the global model, i.e., ${\bf w}^{t+1}= {\bf w}^t+\textrm{CC}(\bar{\bf g}_1^t, \cdots, \bar{\bf g}^t_{\left \lceil  n/s\right \rceil})$.

% \vspace{-3mm}
\section{Threat Model}
\label{sec:threatmodel}
In this section, we discuss the threat model of attacking the SOTA robust AGRs, i.e., targeted poisoning attacks to FLAME, MDAM, {and FLDetector}, and untargeted poisoning attacks to CC and CC-B. Note here that the targeted attack mainly refers backdoor attacks. 

\vspace{+0.05in}
\noindent \textbf{Adversary’s Objective.} 
\emph{For targeted poisoning attacks}, 
an adversary aims to {optimize the existing malicious gradients} obtained by backdoor poisoning  
to evade filtering or/and clipping of the robust AGR (e.g., FLAME, MDAM, and {FLDetector}), so that the resulting global model can achieve a high level on both the main task accuracy and backdoor accuracy. \emph{For untargeted poisoning attacks}, the adversary's goal is to {craft malicious gradients based on benign gradients} to disrupt the server's robust aggregation (e.g., via CC and CC with Bucketing), consequently diminishing the overall main task accuracy of the global model. 

\vspace{+0.05in}
\noindent \textbf{Adversary’s Capability.} 
We assume the total number of malicious clients is $f< n/2$. All malicious clients can collude with each other and have indexes within $[f]$, i.e., from 1 to $f$, without loss of generality. 
For untargeted poisoning attacks, a malicious client can carefully modify its normally trained 
gradient 
to be a malicious one such that it can 
fool the robust AGR. 
For targeted backdoor attacks, we assume the adversary uses the strong model replacement attack~\citep{bagdasaryan2020backdoor}, where 
it aims to replace the true global model ${\bf w}^{t+1}={\bf w}^t+\frac{\eta}{n} \sum_{i=1}^{n}\mathbf{g}^t_{i}$ 
with any model ${\bf x}$ by poisoning the gradients $\mathbf{g}^t_{i\in[f]}$. 
In our scenario, we extend the model replacement attack from 1 to $f$ malicious clients. Specifically, the $f$ poisoned client gradients, denoted as $\mathbf{g}^p_{i \in [f]}$\footnote{For notation simplicity, we will omit the round $t$ in the gradients ${\bf g}_i^t$, and the meaning is clear from  context.} to differentiate with the benign gradients ${\bf g}_{j \in [f+1,n]}$, has the relationship: 
$\sum_{i=1}^{f}\mathbf{g}^p_i =\frac{n}{\eta}({\bf x}-{\bf w}^t)-\sum_{j=f+1}^{n}\mathbf{g}_j$. 
As the global model converges, each local model may be close enough to the global model such that the benign gradients  
start to cancel out, i.e., $\sum_{j=f+1}^{n}\mathbf{g}_j=0$~\citep{bagdasaryan2020backdoor}. 
Hence, we have $\sum_{i=1}^{f}\mathbf{g}^p_i \approx \frac{n}{\eta}({\bf x}-{\bf w}^t)$. Then an adversary can simply solve for the poisoned gradients of the malicious  clients as: $\mathbf{g}^p_{i}\approx \frac{n}{f \eta} ({\bf x}-{\bf w}^t), \forall i \in [f].$

\begin{table}
\renewcommand{\arraystretch}{1.0}
\addtolength{\tabcolsep}{1.1pt}
\caption{The knowledge of adversary. {Note that there exists no tailored attack on FLDetector. Our experimental results in Section~\ref{experiment-CC} show our AGR-agnostic attacks are already highly effective, and they can perform better with more knowledge.}}
% \vspace{-2mm}
\centering
    \begin{tabular}{|c|c|c|}
    \hline
    The SOTA AGRs & \multicolumn{2}{c|}{Adversary’s Knowledge} \\
    \cline{2-3}
    \emph{-our attack} &  AGR tailored & Gradients known \\  
    \hline
    \hline
    FLAME/MDAM & \ding{51}  &  \ding{51} \\
    \cline{2-3}
    \emph{-targeted} & \ding{55}  &  \ding{51} \\
    %\hline
    \cline{2-3}
    CC/CC-B & \ding{51}  &  \ding{55} \\
    \cline{2-3}
    \emph{-untargeted}& \ding{55}  &  \ding{55} \\
    \hline
    {FLDetector\emph{-targeted}} & \ding{55}  &  \ding{55} \\
    \hline
    \end{tabular}
    \vspace{-2mm}
    \label{Attack-Knowledge}
\end{table}

\vspace{+0.05in}
\noindent \textbf{Adversary’s Knowledge.}
We consider two dimensions:  
knowledge of the AGR aggregator and knowledge of the gradients shared by benign clients (see Table \ref{Attack-Knowledge}). According to whether the adversary knows the AGR aggregator, we classify the malicious attacks into \emph{AGR-tailored} and \emph{AGR-agnostic}. Similarly, we also divide the attacks into \emph{gradients-known} and \emph{gradients-unknown} based on whether the adversary is aware of the benign clients' gradients.
Note that the adversary performing the gradients-unknown attack can utilize the clean data of the malicious clients to obtain benign gradients. 

\section{Optimization-Based Poisoning Attacks to SOTA Robust AGRs in FL}
\label{sec:attack}

\subsection{Overview}
{
Recall that SOTA defenses against untargeted and targeted poisoning attacks all design robust AGRs that involve {clipping} or/and {filtering} malicious gradients based on their statistic differences with benign gradients. 
Hence, the main insight of our attack framework is to carefully create malicious gradients $\{\mathbf{g}^c_{i}\}_{i\in [f]}$ to evade these 
operations in SOTA AGRs. Formally, we introduce a general optimization formula as follows:}
\begin{equation} 
\begin{aligned} 
     & \gamma = \underset{\gamma,i \in [f], j\in [f+1,n]}
     {\arg\max}\ 
     dist(\mathbf{g}^c_{i},\mathbf{g}_j), \\ 
      &  s.t.\ \     
      dist(\mathbf{g}^c_{i},\mathbf{g}_j)
     \le th, \, \forall i \in [f], j\in [f+1,n],\\
     & \quad \ \ \ 
     \mathbf{g}^c_{i}=\mathcal{F}(\mathbf{g}^p_{i},\mathbf{g}^b,\gamma), \, \forall i \in [f]. 
\end{aligned} 
\label{equation: overview-general_formula}
\end{equation} 
At a high-level, our objective function aims to create the malicious gradients $\{\mathbf{g}^c_{i}\}_{i\in[f]}$({For ease of description, we also use $\mathbf{g}^c_{i}$ to indicate malicious momentum $\mathbf{m}^c_{i}$}) such that 
their maximum distance from the benign gradients $\{\mathbf{g}_j\}_{j\in[f+1,n]}$ 
is within a predefined/calculated threshold $th$ used by the 
{filtering} or/and {clipping} operation in SOTA AGRs. Note that directly computing $\{\mathbf{g}^c_{i}\}_{i\in[f]}$ is challenging. To address it, we observe malicious gradients can be built (characterized by a function $\mathcal{F}$) from known poisoned gradients $\mathbf{g}^p_{i}$ and a {reference} benign gradient $\mathbf{g}^b$, 
which are coupled with a scaling hyperparameter $\gamma$. 
Hence, the final optimization problem of learning $\{\mathbf{g}^c_{i}\}_{i\in[f]}$  can be reduced to learning the $\gamma$.

Next, we investigate the vulnerabilities of the SOTA robust AGRs (i.e., FLAME, MDAM, {FLDetector}, and CC (with Bucketing)) and then 
instantiate our attack framework in Equation (\ref{equation: overview-general_formula}) by designing optimization-based 
attacks against these AGRs one-by-one. 

\subsection{Targeted Poisoning Attack to FLAME, MDAM, {FLDetector}}
\label{sec:targetedattack}

\noindent {\bf AGR-Tailored Attack to FLAME.} 
As stated in Section~\ref{sec:SOTAAGEs}, 
FLAME first  
limits the gradients with large length to not exceed a clipping threshold $q^t$ in each $t$-th round. To attack this clipping, we deploy projected gradient descent (PGD)  
with the adversary periodically projecting their client gradients on a small ball centered around the global model ${\bf w}^t$. To be specific, the malicious client calculates $q^t$ so that the malicious gradients $\mathbf{g}_{i \in [f]}^p$ respect  
the constraint $ \left \| \mathbf{g}_i^p  \right \|_2 \le \left | q^t- \|{\bf w}^t\|_2 \right |, \forall i \in [f]$, i.e., malicious gradients distribute over a ball with ${\bf w}^t$ as the center and $\left | q^t-\|{\bf w}^t\|_2 \right |$ as the radius.

Additionally, FLAME uses pairwise cosine distances to measure the angular differences between all pairs of model gradients, and applies the HDBSCAN clustering algorithm to further filter the malicious gradients with large angular deviations. 
Hence, a successful attack also requires crafting the malicious gradients (denoted as $\mathbf{g}^c_{i \in [f]}$) to evade the filtering. 
Here, we propose to rotate $\mathbf{g}^p_{i \in [f]} $ to be the corresponding $\mathbf{g}^c_{i \in [f]}$ (with a reference benign gradient $ \mathbf{g}^b$), so that its maximum cosine distance with all benign gradients is not greater than the maximum cosine distance among benign gradients. Formally, 
we can express 
the optimization problem as: 
\begin{equation} 
\begin{aligned} 
     & \gamma = \underset{\gamma,i \in [f], j\in [f+1,n]}
     {\arg\max}\ 
     cos(\mathbf{g}^c_{i},\mathbf{g}_j), \\ 
     &  s.t.\ \  cos(\mathbf{g}^c_{i},\mathbf{g}_j)
     \le \underset{k,j\in [f+1,n]}{\max}\ cos(\mathbf{g}_k,\mathbf{g}_j),  \\ 
     &\quad \ \ \  \mathbf{g}^c_{i}={\| \mathbf{g}^p_{i}\|_2 } / {\| \mathbf{g}^{u}_{i} \|_2 } \cdot \mathbf{g}^{u}_{i};  \ \ 
    \mathbf{g}^{u}_{i}= \mathbf{g}^p_{i}+\gamma (\mathbf{g}^b-\mathbf{g}^p_{i}),
\end{aligned} 
\label{FLAME AGR-Tailored}
%\vspace{-3.3mm} 
\end{equation} 
where we set the reference $\mathbf{g}^b$ as averaging certain benign gradients. 
Specifically, if the benign clients' gradients are known to the adversary (i.e., gradients-known), it can average benign clients' benign gradients, i.e., $\mathbf{g}^b = \textrm{Avg}(\mathbf{g}_{\{ i \in [f+1,n]\}})$. 
On the other hand, when the benign clients' gradients are unknown (i.e., gradients-unknown), the adversary can  average the ``benign'' gradients obtained by  malicious clients on their clean data (i.e., without poisoning), i.e., $\mathbf{g}^b = \textrm{Avg}(\mathbf{g}_{\{ i \in [1,f]\}})$. 
$\gamma$ is the scaling hyperparameter that we aim to learn and ${\bf g}^p_{i \in [f]}$ are defined in Section~\ref{sec:threatmodel}. 

\vspace{+0.05in}
\noindent {\bf AGR-Tailored Attack to MDAM.}
MDAM chooses a set of $n-f$ momentums with the smallest diameter aiming to filter out the possible $f$ malicious clients. 
To break this robust ARG, we need to force it to choose our tailored 
malicious momentums and mix them with benign ones during training. 
Here, we would like to replace (any) $f$ benign clients' momentums that would have been chosen by MDAM with $f$ \emph{identical} tailored malicious momentums.
To be specific, we learn to optimize the 
malicious momentums ${\bf m}^p_{i \in [f]}$ to linearly approach a benign ${\bf m}^b$  until there exists a malicious momentum from ${\bf m}^c_{i \in [f]}$ whose largest distance with respect to the benign momentums in any chosen benign set $S \subset [f+1, n]$ with  $|S|=n-2f $ is less than the maximum distance between any two benign momentums. 
The optimization problem is defined as:
\begin{equation}
\vspace{-0.4mm}
\begin{aligned}  
     & \gamma = \underset{\gamma,i \in [f], j \in S \subset  [f+1, n] \atop |S|=n-2f }{\arg\max} \left \|   {\bf m}^c_{i}-{\bf m}_j\right \|_2,  \\
     & s.t. \ \ \left \|   {\bf m}^c_{i}-{\bf m}_j\right \|_2 \le \underset{k,j\in [f+1,n]}{\max} \left \|  {\bf m}_k-{\bf m}_j \right \|_2, \\    
     &\quad \ \ \ {\bf m}^c_{i}={\bf m}^p_{i}+\gamma ({\bf m}^b-{\bf m}^p_{i}), %\nonumber  
% \vspace{-1mm} 
\end{aligned} 
\label{MDA-ATTACK-MOTHED}
\end{equation}
{where ${\bf m}^b$ is the average of certain benign momentum. Similarly, 
for gradients-known and -unknown, we set ${\bf m}^b = \textrm{Avg}({\bf m}_{\{ i \in [f+1,n]\}})$ and ${\bf m}^b = \textrm{Avg}({\bf m}_{\{ i \in [1,f]\}})$, respectively. 
${\bf m}_{i\in [f]}^p $ are the momentum of malicious gradients ${\bf g}_{i\in [f]}^p$ defined in Section~\ref{sec:threatmodel}, i.e.,  $\mathbf{m}^p_{i}=\beta \mathbf{m}^p_{i}+(1-\beta)\mathbf{g}^p_{i}, \forall i \in [f] $}.

\vspace{+0.05in}
\noindent {\bf AGR-Agnostic Attacks (to FLAME, MDAM, and FLDetector).} 
\emph{It is challenging to design tailored attacks on FLDetector, due to it uses the history information in each round that is not easy to incorporate into the attack optimization. To relax it, we propose an AGR-agnostic attack formulation; and 
if the AGR-agnostic attack is already effective, then AGR-tailored attack can performance better.
}
Specifically, in this setting, the adversary does not know which AGR the defense uses. 
To design effective AGR-agnostic attacks, we need to uncover the shared property in the existing robust AGRs. 
Particularly, we note that, though with different techniques, 
existing robust AGRs mainly perform statistical analysis on client models and 
identify malicious clients as those largely deviate from others based on some similarity metric such as Euclidean distance and cosine similarity. 
Based on this intuition, we hence propose to adjust the malicious gradients to be close to the benign ones such that these malicious gradients can be selected by (any) robust AGRs. 
W.l.o.g, we use Euclidean distance as the similarity metric 
and extend Equation (\ref{MDA-ATTACK-MOTHED}) to a more general case. Specifically, we optimize $\mathbf{g}^p_{i \in [f]} $ to approach $\mathbf{g}^b$ so that its maximum distance with respect to any benign gradient is upper bounded by the maximum distance between any two benign gradients. Formally, the equation can be expressed as follows: 
\begin{equation}
\begin{aligned}
% \small
     & \gamma = \underset{\gamma, i \in [f], j\in [f+1,n]}{\arg\max} \left \| \mathbf{g}^c_{i}-\mathbf{g}_j\right \|_2, \\ \label{AGR-Agnostic}
     & s.t. \ \ \left \| \mathbf{g}^c_{i}- \mathbf{g}_j\right \|_2   \le  \underset{k,j\in [f+1,n]}{\max}\ \left \|  \mathbf{g}_k-\mathbf{g}_j \right \|_2,  \\
     & \quad \ \ \  \mathbf{g}^c_{i}=\mathbf{g}^p_{i}+\gamma (\mathbf{g}^b-  \mathbf{g}^p_{i}). %\nonumber
\end{aligned} 
\end{equation}
\subsection{Untargeted Poisoning Attack to CC and CC with Bucketing}
\label{sec:untargetedattack}

%\vspace{+0.05in}
\noindent {\bf AGR-Tailored Attack to CC.}
\label{4.2CC}
CC clips each client gradient $\mathbf{g}_{i}$ by $\mathbf{g}_i\min (1,\frac{\tau}{\left \| \mathbf{g}_i\right \|_2 } )$ 
to 
reduce the negative effect caused by malicious gradients with a large length. 
However, we observe that clipping is ineffective if we can always maintain $\frac{\tau}{\left \| \mathbf{g}_i\right \|_2 } \geq 1$, 
i.e., $\left \| \mathbf{g}_{i}  \right \|_2 \le \tau$. Under this, we can create the malicious gradients $\mathbf{g}_{i \in [f]}^c$ such that: 
\begin{equation} %\vspace{-2mm}
\begin{aligned}
    & \gamma = \underset{\gamma, i \in [f]}{\arg\max} 
    \left \| \mathbf{g}_{i}^c  \right \|_2, \\
    & s.t.  \left \| \mathbf{g}_{i}^c  \right \|_2 \le \tau,  \ \
    \mathbf{g}_{i}^c =  \mathbf{g}^b+\gamma{\mathbf{g}^p_{i}}, \ \ 
    \mathbf{g}^p_{i}=-\textrm{sign}(\mathbf{g}^b). \ 
\end{aligned}
\label{attack-CC}
%\vspace{-2mm} 
\end{equation} 
There are many ways to set malicious gradients 
$\mathbf{g}^p_{i \in [f]}$. 
Motivated by \citep{shejwalkar2021manipulating}, one simple yet effective way is setting $\mathbf{g}^p_{i \in [f]}$ as the \emph{inverse} sign of the reference benign gradient $\mathbf{g}^b$, which can be a  randomly chosen benign gradient. 

\vspace{+0.05in}
\noindent {\bf AGR-Tailored Attack to CC-B.}
\label{4.3Bucketing}
Our attack on CC with bucketing is similar to that on CC. 
As seen in Section~\ref{sec:SOTAAGEs}, bucketing is a post-processing step on client gradients mainly mitigating the non-IID data across clients.  
In principle, a successful attack on CC will render the bucketing aggregation ineffective as well. Here, we 
directly craft malicious gradients 
using Equation (\ref{attack-CC}) and then perform CC aggregation with bucketing on the crafted malicious gradients.

\vspace{+0.05in}
\noindent {\bf AGR-Agnostic Attack (to CC and CC-B).} $\tau$ is the only hyperparameter in CC. In this attack setting, $\tau$ is unknown to the adversary. 
To differentiate between $\tau$ used in CC and that in the attack, we name it as  CC-$\tau$ and ATK-$\tau$, respectively. 
In practice, the used CC-$\tau$'s often have a format of $10^x$. 
Hence the adversary can exploit these potential values as ATK-$\tau$ to execute the attack in Equation (\ref{attack-CC}). Note that there is no such limitation and our experimental results in Section \ref{experiment-CC}  show a larger value of ATK-$\tau$ always ensures the attack to be effective, whatever the true CC-$\tau$ is.

%\vspace{-0.8em}
\subsection{Solving for The Scaling Hyperparameter $\gamma $}
\label{sec:gamma}
\subsubsection{Binary Search Algorithm}
%\noindent {\bf {Solving by Algorithm~\ref{alg:attack}.}} 
Both the targeted and untargeted poisoning attacks to FL involve optimizing the scaling hyperparameter $\gamma$. For the untargeted attacks,  
the malicious gradients $\mathbf{g}^c_{i \in [f]}$ are obtained by \emph{maximizing} 
$\gamma$ to render the attack $\mathbf{g}_{i \in [f]}^c=\mathbf{g}^b+\gamma \mathbf{g}^p_{i \in [f]}$ be effective. 
For the targeted attacks, 
in contrast, 
an adversary generates $\mathbf{g}^c_{i \in [f]}$ 
by exploring the \emph{minimum} $\gamma$ to maintain the attack effectiveness, e.g., $\mathbf{g}^c_{i \in [f]}=\mathbf{g}^p_{i \in [f]}+\gamma (\mathbf{g}^b-\mathbf{g}^p_{i \in [f]})$. 
Algorithm~\ref{alg:attack} shows the details of solving $\gamma$ in the two  scenarios. 
Specifically, 
we start with a small (or large) $\gamma$ for untargeted (or targeted) attacks and increase (or decrease) $\gamma$ with a step size \textit{step} until $\mathcal{O}$ returns ``True'', 
where $\mathcal{O}$ takes as input the union of the  malicious and benign gradients, i.e., $\mathbf{g}_{i \in [n]}= \mathbf{g}^c_{i \in [f]} \cup 
\mathbf{g}_{\{ i \in [f+1,n]\}}$, and $\gamma$, and outputs ``True'' if the obtained $\mathbf{g}^c_{i \in [f]}$ in Equations (\ref{FLAME AGR-Tailored})-(\ref{attack-CC}) satisfies the adversarial objective. We halve the step size for each $\gamma$ update to make the search finer. 

\vspace{+0.05in}
\noindent{{\bf Complexity Analysis.}} Algorithm~\ref{alg:attack} iteratively optimizes $\gamma$ to satisfy the condition $\mathcal{O}(\mathbf{g}_{i \in [n]},\gamma) ==$ True. In each iteration, it verifies the adversarial objective in Equations (\ref{FLAME AGR-Tailored})-(\ref{attack-CC}), e.g., computes the pairwise gradient/momentum distance among benign clients, and that between malicious clients and benign clients and checks whether the inequality satisfies. The complexity per iteration is O($(n^2-f^2)* \#$model parameters), where $n$ and $f$ are the total number of clients and malicious clients, respectively. On the other hand, 
Algorithm~\ref{alg:attack} guarantees to converge and stops when $\text{step} \leq \varepsilon$. {Note that $\text{step}$ is initialized as $\gamma_{init}/2$ and halved in each iteration. So the convergence iteration $m$ is obtained when $\gamma_{init} / 2^{m+1} \leq \varepsilon$, which means $m = \log_2 (\gamma_{init} / \varepsilon)-1$.}

\begin{algorithm}[!t]
\small
\caption{Learning the Scaling Hyperparameter $\gamma$} 
\begin{flushleft}
{\bf Input:} $\gamma_{init}, \varepsilon, \mathcal{O}, \mathbf{g}_{i \in [n]}$

{\bf Output: $\gamma_{succ}$}
\end{flushleft}
\begin{algorithmic}[1]
\State step $ \gets \gamma_{init}/2, \gamma \gets \gamma_{init}$ 	
\While{$\left |\gamma_{succ} - \gamma \right | > \varepsilon $ } 
    \If{$\mathcal{O}(\mathbf{g}_{i \in [n]},\gamma)==$ True} 
        \State $\gamma_{succ}\gets \gamma$ 
        \If{ attacking FLAME, MDAM, or FLDetector}  
            \State $\gamma \gets (\gamma-$step$/2)$ 
        \Else
            \State $\gamma \gets (\gamma+$step$/2)$ 
        \EndIf
    \Else
        \If{ attacking FLAME, MDAM, or FLDetector}   
            \State $\gamma \gets (\gamma+$step$/2)$ 
        \Else
            \State $\gamma \gets (\gamma-$step$/2)$
        \EndIf
    \EndIf
    \State step=step$/2$
\EndWhile
\end{algorithmic}
\label{alg:attack}
\vspace{-0.1em} 
\end{algorithm} 
\setlength{\textfloatsep}{+4mm}

\vspace{+0.05in}
\subsubsection {Analytic Solution} We notice that $\gamma$ can be computed directly with an analytic solution. Yet this approach incurs higher computational costs than our iterative solver. Particularly, the direct solver has a total complexity of $24 * \#$model parameters, while our iterative solver has a complexity of $4* \#$model parameters per iterations. 
As the number of iterations (controlled by $\gamma_{init}$ and $\epsilon$) is small ($\le 5$ in our experiments), the total complexity of our iterative solver is slightly smaller than the direct solver. 

\begin{table*}[!t]
\renewcommand{\arraystretch}{0.9}
\addtolength{\tabcolsep}{6pt}
\caption{Results of our attack and the SOTA DBA against FLAME under various threat models. 
{Our attacks show an BA improvement ranging from 31\% to 94\% over the SOTA DBA with comparable or better MA under the same settings.}
}
%\vspace{-2mm}
\centering
    \begin{tabular}{|c|c|c|c|c|c|c|c|}
    \hline
    & & & & \multicolumn{2}{c|}{Gradients known} &\multicolumn{2}{c|}{Gradients unknown} \\
    \cline{5-8}
    %\hline
    Dataset  & No attack & $f/n$ & DBA & AGR & AGR & AGR & AGR \\ % 
    %\cline{5-8}
    %\hline
    &(MA) & ($\%$) & (MA\ /\ BA) & tailored &  agnostic & tailored &  agnostic \\
    \hline
     &  & 2 & 0.85\ /\ 0.02 & 0.93\ /\ {0.61}  & 0.91\ /\ 0.53 & 0.90\ /\ {0.57} & 0.90\ /\ 0.43\\
     \cline{3-8}
    FMNIST & 0.92  & 5 &  0.88\ /\ 0.01 & 0.91\ /\ {0.84} & 0.90\ /\ 0.71 & 0.91\ /\ {0.77} & 0.89\ /\ 0.66\\
    \cline{3-8}
     &   & 10 &  0.85\ /\ 0.01 & 0.92\ /\ {0.95} & 0.92\ /\ 0.83 &0.92\ /\ {0.90} & 0.92\ /\ 0.77\\
     \cline{3-8}
     & & 20 & 0.86\ /\ 0.19 & 0.92\ /\ {0.96} & 0.90\ /\ 0.84 & 0.92\ /\ {0.93} & 0.92\ /\ 0.74\\
    \hline
     &  & 2 & 0.70\ /\ 0.03 & 0.72\ /\ {0.65} & 0.71\ /\ 0.60 & 0.69\ /\ {0.53} & 0.70\ /\ 0.48\\
     \cline{3-8}
    CIFAR10 & 0.71  & 5 & 0.70\ /\ 0.13 & 0.70\ /\ 0.75 & 0.72\ /\ {0.74} & 0.70\ /\ {0.74} & 0.72\ /\ 0.50\\
    \cline{3-8}
     &  & 10 &  0.67\ /\ 0.18 & 0.72\ /\ 0.79 & 0.70\ /\ {0.78} & 0.67 /\ {0.79} & 0.68\ /\ 0.54 \\
     \cline{3-8}
     &  & 20 & 0.69\ /\ 0.27 & 0.72\ /\ {0.79} & 0.72\ /\ 0.79 & 0.72\ /\ {0.80} & 0.72\ /\ 0.63\\
     \hline
      &  & 2 & 0.90\ /\ 0.20 &  0.91\ /\ {0.64} &  0.91\ /\ 0.58 &  0.92\ /\ {0.60} & 0.91\ /\ 0.51\\
      \cline{3-8}
     FEMNIST & 0.94 & 5 & 0.89\ /\ 0.19 &  0.90\ /\ {0.89} & 0.92\ /\ 0.82 &  0.90\ /\ {0.82} & 0.90\ /\ 0.74\\
     \cline{3-8}
      &  & 10 &  0.90\ /\ 0.31 &  0.91\ /\ {0.92} & 0.90\ /\ 0.83 &  0.91\ /\ {0.86} & 0.92\ /\ 0.82\\
      \cline{3-8}
      &  & 20 &  0.91\ /\ 0.43 &  0.91\ /\ {0.93} & 0.92\ /\ 0.91 &  0.92\ /\ {0.92} & 0.91\ /\ 0.86\\
    \hline
    \end{tabular}
    \label{FLAME-results}  
\vspace{-2mm}
\end{table*}

\begin{table*}[!t]
%\footnotesize
\renewcommand{\arraystretch}{0.9}
\addtolength{\tabcolsep}{5pt}
\caption{Results of our attack and %the SOTA 
DBA against MDAM {under various threat models with the momentum coefficient $\beta=0.9$}. Our attacks significantly outperforms DBA under the same setting. 
{For instance, when $f/n \geq 20\%$, our attacks show an BA improvement ranging from 6\% to 98\% over the SOTA DBA with comparable MA.}
}
\vspace{-2mm}
\centering
    \begin{tabular}{|c|c|c|c|c|c|c|c|}
    % \footnotesize
    \hline
    & & & & \multicolumn{2}{c|}{Gradients known} &\multicolumn{2}{c|}{Gradients unknown} \\
    \cline{5-8}
    %\hline
    Dataset  & No attack & $f/n$ & DBA & AGR & AGR & AGR & AGR \\ % 
    %\cline{5-8}
    %\hline
    &(MA) & ($\%$) & (MA\ /\ BA) & tailored &  agnostic & tailored &  agnostic \\
    \hline
     &  & 5 &  0.93\ /\ 0.01  &  0.93\ /\ 0.01   &  0.93\ /\ 0.01  &  0.92\ /\ 0.01  &  0.92\ /\ 0.01 \\
     \cline{3-8}
    FMNIST & 0.93  & 10 &  0.93\ /\ 0.01  &  0.93\ /\ 0.01 &  0.92\ /\ 0.01  &  0.93\ /\ 0.01  &  0.93\ /\ 0.01 \\
    \cline{3-8}
     &   & 20 & 0.93\ /\ 0.01  &  0.93\ /\ 0.99   &  0.92\ /\ 0.61  &  0.92\ /\ 0.68  &  0.93\ /\ 0.45 \\
     \cline{3-8}
     & & 30 &  0.93\ /\ 0.01  &  0.93\ /\ 1.00  &  0.92\ /\ 0.74  &  0.91\ /\ 0.99  &  0.92\ /\ 0.67 \\
    \hline
     &  & 5 &  0.73\ /\ 0.03  &  0.73\ /\ 0.14   &  0.73\ /\ 0.11  &  0.73\ /\ 0.11  &  0.73\ /\ 0.08 \\
     \cline{3-8}
    CIFAR10 & 0.73  & 10 &  0.73\ /\ 0.03  &  0.73\ /\ 0.67   &  0.73\ /\ 0.56  &  0.73\ /\ 0.46  &  0.72\ /\ 0.43 \\
    \cline{3-8}
     &  & 20 &  0.73\ /\ 0.05  &  0.72\ /\ 0.81   &  0.73\ /\ 0.81  &  0.74\ /\ 0.78   &  0.73\ /\ 0.80 \\
     \cline{3-8}
     &  & 30 &  0.73\ /\ 0.47  &  0.73\ /\ 0.92   &  0.72\ /\ 0.89  &  0.73\ /\ 0.91  &  0.72\ /\ 0.88 \\
     \hline
      &  & 5 & 0.96\ /\ 0.24  &  0.96\ /\ 0.30   &  0.96\ /\ 0.27  &  0.96\ /\ 0.29  &  0.96\ /\ 0.24 \\
      \cline{3-8}
     FEMNIST & 0.96 & 10 & 0.95\ /\ 0.43  &  0.96\ /\ 0.83   &  0.96\ /\ 0.64  &  0.95\ /\ 0.44  &  0.95\ /\ 0.47 \\
     \cline{3-8}
      &  & 20 &  0.96\ /\ 0.68  &  0.96\ /\ 0.89   &  0.96\ /\ 0.88  &  0.96\ /\ 0.74  &  0.95\ /\ 0.78 \\
      \cline{3-8}
      &  & 30 &  0.94\ /\ 0.76  &  0.95\ /\ 0.96   &  0.95\ /\ 0.95  &  0.94\ /\ 0.87  &  0.95\ /\ 0.85 \\
    \hline
    \end{tabular}
    \label{MDAM-results}
 \vspace{-2mm}   
\end{table*}
%--------------------------------------------------
\begin{table}[!t]
\renewcommand{\arraystretch}{0.9}
\addtolength{\tabcolsep}{-2pt}
\caption{{Results of our AGR-agnostic and gradient-unknown targeted poisoning attack against FLDetector.} The ``No attack (MA)" are 0.93, 0.73, and 0.96 on the FMNIST, CIFAR10, and FEMNIST datasets, respectively. Note that FLDetector is customized for defending against the centralized BA.}
%\vspace{-2mm}
\centering
    \begin{tabular}{|c|c|c|c|c|}
    \hline 
    \multirow{2}*{\diagbox{Dataset}{$f/n$ ($\%$)}} & 5 & 10 & 20 & 30 \\
    \cline{2-5}
     & MA\ /\ BA & MA\ /\ BA & MA\ /\ BA & MA\ /\ BA \\ 
    \hline
    FMNIST & 0.93\ /\ 0.86 & 0.92\ /\ 0.99 & 0.93\ /\ 1.00 & 0.93\ /\ 0.99 \\
    \hline
    CIFAR10 & 0.75\ /\ 0.25 & 0.74\ /\ 0.72 & 0.75\ /\ 0.84 & 0.74\ /\ 0.89 \\
    \hline
    FEMNIST & 0.90\ /\ 0.48 & 0.90\ /\ 0.63 & 0.90\ /\ 0.73 & 0.90\ /\ 0.90 \\
    \hline
    \end{tabular}
    \label{FLDetector-results}
    % \vspace{-2mm}
\end{table}

\noindent{{\bf Complexity Analysis.}} 
{Taking Equation (4) for instance,} it can be rewritten as a quadratic inequation with one unknown variable $\gamma$: 
$$\underset{j\in [f+1,n]}\max || \mathbf{g}^p_{i} - \mathbf{g}_j   +\gamma (\mathbf{g}^b -\mathbf{g}^p_i) ||_2 \le \underset{k,j\in [f+1,n]}{\max}\ || \mathbf{g}_k-\mathbf{g}_j ||_2.  \quad (6) $$
To solve Equation (6), we need to compute $\gamma$ for each $j \in [f+1, n]$, store the maximum value of the LHS term w.r.t. $\forall j \in [f+1, n]$, and verify whether it is less than or equal to the RHS term. For brevity, we simply consider a certain $j$.   
Denote the RHS term as 
$c = \underset{k,j \in [f+1,n]}{\max}\ || \mathbf{g}_k-\mathbf{g}_j ||_2$,  
which can be calculated once and stored (and has constant time).  
Then the inequality becomes 
$A_j \gamma^2 + B_j \gamma + C \leq 0 $, 
where $A_j = || \mathbf{g}^p_{i} - \mathbf{g}_j ||_2^2, \, B_j = 2 \langle \mathbf{g}^p_{i} - \mathbf{g}_j,  \mathbf{g}^b -\mathbf{g}^p_i \rangle, \, C = || \mathbf{g}^b -\mathbf{g}^p_i ||^2 - c$. 

Calculating $A_j, B_j$ and $C$ all have the complexity $4d$, where $d$ is the number of model parameters, so the total complexity is $12d$.   
The optimal solution $\gamma$ is from the two roots: $\gamma = \big(-B_j \pm  \sqrt{B_j^2 - 4 A_j C}\big) /{2 A_j}$. 
Since there are two possible solutions of $\gamma$, we need to check each of the two roots and decide the optimal $\gamma$ that yields the largest value of the LHS. So the total complexity is $24d$.  
In our attack, each iteration calculates LHS of Equation (6), which has complexity $4d$. As the number of iterations (controlled by $\gamma_{init}$ and $\epsilon$) is small (i.e., 4-5 in our experiments), the total complexity of our iterative attack is slightly smaller.

\section{Experiments}
\label{experiment-CC}
In this section, we evaluate our attack {framework} on the SOTA robust AGRs. 
We first set up the experiments and then show the attack results on each robust AGR. 

% \vspace{-1em}
\subsection{Experimental Setup}
\noindent {\bf Datasets and Architectures.} 
{Following existing works~\citep{farhadkhani2022byzantine,karimireddy2021learning,zhang2022fldetector,ozdayi2021defending,xie2020dba}}, 
we consider FL on three benchmark image datasets, i.e., FMNIST, CIFAR10, and FEMNIST~\citep{caldas2018leaf}, where the first two datasets are IID distributed, while  the third one is non-IID distributed. 
{FMNIST has 60K training images and 10K testing images from 10 classes with image size 28x28. CIFAR10 has 50K training images and 10K testing images. FEMNIST includes 3,383 clients, 62 classes, and a total of 805,263 grayscale images. For FL training, we selected 1,000 out of the 3,383 clients.} 
For FMNIST and FEMNIST, we consider a convolutional neural network (CNN) with 2 convolutional (Conv) layers followed by 2 fully-connected (FC) layers. While for CIFAR10, we use a CNN with 3 Conv layers and 3 FC layers on the targeted attack, and a ResNet-20 \citep{he2016deep} on the untargeted attack. 
We follow FLAME, MDAM, {FLDetector}, CC, CC-B to set the number of total clients and fraction of clients selected in each round. For instance, the total number of clients in CC is 50, and all clients are selected for training. {We set the backdoor classes to $4$ and $7$ for targeted poisoning attacks, and} 
use the cross entropy loss for federated training. We set the learning rate for server's global aggregation as $1.0$ and client training as $0.1$ and total number of rounds is 200. 

\begin{table*}[!t]
%\footnotesize
\renewcommand{\arraystretch}{0.9}
\addtolength{\tabcolsep}{1.1pt}
\caption{Results of our attack and the SOTA DBA against MDAM {under various threat models with the momentum coefficient $\beta=0$, $0.6$, and $0.99$}. Our attack significantly outperforms DBA when $f/n$ is large. 
}
\centering
    \begin{tabular}{|c|c|c|c|c|c|c|c|c|}
    % \footnotesize
    \hline
    & & & & & \multicolumn{2}{c|}{Gradients known} &\multicolumn{2}{c|}{Gradients unknown} \\
    \cline{6-9}
    %\hline
    $\beta$ & Dataset  & No attack & $f/n$ & DBA & AGR & AGR & AGR & AGR \\ % 
    %\cline{5-8}
    %\hline
    & &(MA) & ($\%$) & (MA/BA) & tailored &  agnostic & tailored &  agnostic \\
    \hline
    & &  & 5  & 0.92\ /\ 0.01 & 0.92\ /\ 0.01 & 0.93\ /\ 0.01 & 0.93\ /\ 0.00 & 0.92\ /\ 0.01 \\ 
    \cline{4-9}
    & FMNIST & 0.93 & 10 & 0.93\ /\ 0.00 & 0.93\ /\ 0.01 & 0.92\ /\ 0.01 & 0.93\ /\ 0.00 & 0.92\ /\ 0.01\\ 
    \cline{4-9}
    & &  & 20 & 0.93\ /\ 0.01 & 0.93\ /\ 0.95 & 0.93\ /\ 0.98 & 0.93\ /\ 0.99 & 0.92\ /\ 0.94\\ 
    \cline{4-9}
    & &  & 30 & 0.93\ /\ 0.01 & 0.93\ /\ 1.00 & 0.92\ /\ 1.00 & 0.92\ /\ 1.00 & 0.92\ /\ 1.00 \\ 
    \cline{2-9}

    & &   & 5 & 0.72\ /\ 0.03 & 0.71\ /\ 0.04 & 0.72\ /\ 0.01 & 0.73\ /\ 0.01  & 0.73\ /\ 0.01\\ 
    \cline{4-9}
    0 & CIFAR10 & 0.72 & 10  & 0.71\ /\ 0.07 & 0.72\ /\ 0.76 & 0.72\ /\ 0.74 &0.73\ /\ 0.71 & 0.74\ /\ 0.69  \\ 
    \cline{4-9}
    & &  & 20 & 0.72\ /\ 0.27 & 0.73\ /\ 0.91 & 0.73\ /\ 0.89  & 0.73\ /\ 0.84  & 0.73\ /\ 0.82\\ 
    \cline{4-9}
    & &  & 30 & 0.73\ /\ 0.27 & 0.73\ /\ 0.96 & 0.73\ /\ 0.91&0.73\ /\ 0.92 & 0.73\ /\ 0.91\\ 
    \cline{2-9}

    & & & 5 & 0.97\ /\ 0.22 & 0.97\ /\ 0.53 & 0.96\ /\ 0.47 & 0.96\ /\ 0.44 & 0.96\ /\ 0.32  \\ 
    \cline{4-9}
    & FEMNIST & 0.97 & 10  & 0.97\ /\ 0.40 & 0.97\ /\ 0.83 & 0.96\ /\ 0.53 & 0.96\ /\ 0.50 & 0.96\ /\ 0.39\\ 
    \cline{4-9}
    & &  & 20 & 0.96\ /\ 0.64 & 0.96\ /\ 0.95 & 0.96\ /\ 0.73 & 0.96\ /\ 0.72 & 0.97\ /\ 0.65\\ 
    \cline{4-9}
    & &  & 30 & 0.94\ /\ 0.78 & 0.97\ /\ 0.99 & 0.96\ /\ 0.96 & 0.96\ /\ 0.96 & 0.96\ /\ 0.88\\ 
    \cline{1-9}
    %--------------------------
    & & & 5 & 0.93\ /\ 0.01 & 0.92\ /\ 0.02 & 0.93\ /\ 0.00 & 0.93\ /\ 0.00 & 0.92\ /\ 0.01\\ 
    \cline{4-9}
    & FMNIST & 0.93 & 10  &  0.93\ /\ 0.00 &  0.92\ /\ 0.02 & 0.93\ /\ 0.01 & 0.93\ /\ 0.01 & 0.93\ /\ 0.01\\ 
    \cline{4-9}
    & &  & 20 & 0.93\ /\ 0.00 & 0.93\ /\ 0.89 & 0.92\ /\ 0.83 & 0.92\ /\ 0.88 & 0.92\ /\ 0.77\\ 
    \cline{4-9}
    & &  & 30 & 0.92\ /\ 0.01 & 0.92\ /\ 0.97 & 0.93\ /\ 0.95 & 0.92\ /\ 0.95 & 0.93\ /\ 0.92\\ 
    \cline{2-9}
    
    & &  & 5 & 0.73\ /\ 0.03 & 0.73\ /\ 0.03 & 0.73\ /\ 0.01 & 0.73\ /\ 0.01 & 0.73\ /\ 0.01 \\ 
    \cline{4-9}
    0.6 & CIFAR10 & 0.73 & 10 & 0.72\ /\ 0.03 & 0.73\ /\ 0.69 & 0.73\ /\ 0.67 & 0.73\ /\ 0.65 & 0.72\ /\ 0.41 \\ 
    \cline{4-9}
    & &  & 20 & 0.72\ /\ 0.05 & 0.73\ /\ 0.85 & 0.73\ /\ 0.84 & 0.73\ /\ 0.84 & 0.72\ /\ 0.77\\ 
    \cline{4-9}
    & &  & 30 & 0.69\ /\ 0.41 & 0.72\ /\ 0.91 &  0.72\ /\ 0.93 & 0.73\ /\ 0.93& 0.73\ /\ 0.88\\ 
    \cline{2-9}

    & & & 5 & 0.96\ /\ 0.29 & 0.96\ /\ 0.54 & 0.96\ /\ 0.47 & 0.96\ /\ 0.47 & 0.96\ /\ 0.45 \\ 
    \cline{4-9}
    & FEMNIST & 0.96 & 10 & 0.95\ /\ 0.59 & 0.96\ /\ 0.76 & 0.96\ /\ 0.70 & 0.96\ /\ 0.65 & 0.96\ /\ 0.63 \\ 
    \cline{4-9}
    & &  & 20 & 0.96\ /\ 0.68 & 0.96\ /\ 0.91 & 0.96\ /\ 0.89 & 0.96\ /\ 0.79 & 0.95\ /\ 0.74\\ 
    \cline{4-9}
    & &  & 30 & 0.94\ /\ 0.79 & 0.95\ /\ 0.98 & 0.96\ /\ 0.93 & 0.96\ /\ 0.95& 0.96\ /\ 0.87 \\ 
    \cline{1-9}
    %------------------------
    & & & 5  & 0.92\ /\ 0.01 & 0.92\ /\ 0.01 & 0.92\ /\ 0.01 & 0.92\ /\ 0.01 & 0.92\ /\ 0.01\\ 
    \cline{4-9}
    & FMNIST & 0.92 & 10  & 0.92\ /\ 0.01 & 0.92\ /\ 0.01 & 0.93\ /\ 0.06 & 0.92\ /\ 0.01 & 0.92\ /\ 0.01\\ 
    \cline{4-9}
    & & & 20   & 0.92\ /\ 0.04 & 0.92\ /\ 0.71 & 0.93\ /\ 0.66 & 0.92\ /\ 0.60 & 0.92\ /\ 0.54\\ 
    \cline{4-9}
    & & & 30   & 0.92\ /\ 0.02 & 0.92\ /\ 1.00 & 0.93\ /\ 0.96 & 0.92\ /\ 0.86 & 0.92\ /\ 0.82\\ 
    \cline{2-9}
    
    & & & 5  & 0.71\ /\ 0.03 & 0.71\ /\ 0.06 & 0.72\ /\ 0.01 & 0.72\ /\ 0.01 & 0.72\ /\ 0.01\\ 
    \cline{4-9}
    0.99 & CIFAR10 & 0.72 & 10  & 0.70\ /\ 0.06 & 0.72\ /\ 0.59 & 0.72\ /\ 0.58 & 0.72\ /\ 0.54 & 0.72\ /\ 0.51\\ 
    \cline{4-9}
    & & & 20  & 0.71\ /\ 0.25  & 0.72\ /\ 0.71 & 0.72\ /\ 0.63 & 0.72\ /\ 0.65 & 0.72\ /\ 0.58\\ 
    \cline{4-9}
    & & & 30  & 0.72\ /\ 0.35 & 0.71\ /\ 0.83 & 0.72\ /\ 0.75 & 0.72\ /\ 0.87 & 0.72\ /\ 0.65\\ 
    \cline{2-9}

    & & & 5  & 0.96\ /\ 0.44 & 0.96\ /\ 0.55 & 0.96\ /\ 0.47  & 0.96\ /\ 0.46 & 0.96\ /\ 0.43\\ 
    \cline{4-9}
    & FEMNIST & 0.96 & 10 & 0.96\ /\ 0.55 & 0.96\ /\ 0.84 & 0.96\ /\ 0.79  & 0.96\ /\ 0.64 & 0.96\ /\ 0.60\\ 
    \cline{4-9}
    & & & 20  & 0.95\ /\ 0.64 & 0.96\ /\ 0.92 & 0.96\ /\ 0.87  & 0.96\ /\ 0.75 & 0.96\ /\ 0.67\\ 
    \cline{4-9}
    & & & 30& 0.93\ /\ 0.78 & 0.96\ /\ 0.97 & 0.96\ /\ 0.94 & 0.95\ /\ 0.84 & 0.96\ /\ 0.79 \\ 
    \hline    
    \end{tabular}
    \label{app:MDAM-results}
 %\vspace{-3mm}   
\end{table*}

\begin{table*}[!t]
%\footnotesize
\renewcommand{\arraystretch}{0.9}
\addtolength{\tabcolsep}{-1.0pt}
\caption{Results of attacking CC on IID FMNIST and CIFAR10, and attacking CC-B on non-IID FEMNIST with CC-$\tau$ = 10. Our attacks significantly outperform AGR-agnostic LIE and AGR-tailored Fang, especially when $f/n$ is large and dataset is non-IID. 
}
% %\vspace{-2mm}
\centering
    \begin{tabular}{|c|c|c|c|c|c|c|c|c|c|c|c|c|c|c|c|c|c|}
    % \footnotesize
    \hline
    & & & & \multicolumn{7}{c|}{Gradients known} &\multicolumn{7}{c|}{Gradients unknown} \\
    \cline{5-18}
    %\hline
    Dataset & No attack & $f/n$ & LIE & \multicolumn{2}{c|}{AGR-tailored} & \multicolumn{5}{c|}{AGR-agnostic (ATK-$\tau$)} & \multicolumn{2}{c|}{AGR-tailored} & \multicolumn{5}{c|}{AGR-agnostic (ATK-$\tau$)} \\ % 
    \cline{5-18} 
    &(MA) & ($\%$) & & Fang & {Ours} & 0.1 & 1 & 10 & 100 & 1000 & Fang & Ours & 0.1 & 1 & 10 & 100 & 1000 \\
    \hline
    & & 2 & 0.84 & 0.76 & 0.74 & 0.84 & 0.82 & 0.74 & 0.75 & 0.76 & 0.76 & 0.75 & 0.84 & 0.83 & 0.75 & 0.74 & 0.74 \\
    \cline{3-18}
    FMNIST & 0.84 & 5 & 0.84 & 0.76 & 0.72 & 0.84 & 0.82 & 0.72 & 0.73 & 0.72 & 0.77 & 0.69 & 0.84 & 0.81 & 0.69 & 0.71 & 0.70 \\
    \cline{3-18}
    & & 10 & 0.84 & 0.65 & 0.58 & 0.84 & 0.78 & 0.58 & 0.59 & 0.65 & 0.69 & 0.62 & 0.84 & 0.77 & 0.62 & 0.67 & 0.48 \\
    \cline{3-18}
    & & 20 & 0.84 & 0.49 & 0.40 & 0.83 & 0.76 & 0.40 & 0.11 & 0.10 & 0.51 & 0.47 & 0.83 & 0.76 & 0.47 & 0.15 & 0.09 \\
    \hline
    %------------------------------------------------
    & & 2 & 0.66 & 0.66 & 0.64 & 0.65 & 0.65 & 0.64 & 0.10 & 0.11 & 0.69 & 0.60 & 0.65 & 0.65 & 0.60 & 0.12 & 0.10 \\
    \cline{3-18}
    CIFAR10 & 0.66 & 5 & 0.60 & 0.66 & 0.45 & 0.65 & 0.63 & 0.45 & 0.32 & 0.19 & 0.66 & 0.48 & 0.65 & 0.65 & 0.48 & 0.47 & 0.21 \\
    \cline{3-18}
    & & 10 & 0.49 & 0.65 & 0.11 & 0.65 & 0.64 & 0.11 & 0.08 & 0.09 & 0.66 & 0.14 & 0.62 & 0.64 & 0.14 & 0.11 & 0.10 \\
    \cline{3-18}
    & & 20 & 0.43 & 0.21 & 0.11 & 0.61 & 0.60 & 0.11 & 0.09 & 0.13 & 0.25 & 0.10 & 0.62 & 0.57 & 0.10 & 0.09 & 0.11\\
    \hline
    %------------------------------------------------
    & & 2 & 0.92 & 0.89 & 0.83 & 0.92 & 0.91 & 0.83 & 0.11 & 0.11 & 0.90 & 0.87 & 0.92 & 0.90 & 0.87 & 0.12 & 0.09  \\
    \cline{3-18}
    FEMNIST & 0.92 & 5 & 0.89 & 0.80 & 0.10 & 0.92 & 0.90 & 0.10 & 0.09 & 0.11 & 0.88 & 0.09 & 0.92 & 0.89 & 0.09 & 0.12 & 0.09  \\
    \cline{3-18}
    & & 10 & 0.79 & 0.65 & 0.09 & 0.92 & 0.88 & 0.09 & 0.10 & 0.09 & 0.80 & 0.12 & 0.92 & 0.87 & 0.12 & 0.09 & 0.11 \\
    \cline{3-18}
    & & 20 & 0.57 & 0.32 & 0.09 & 0.91 & 0.86 & 0.09 & 0.07 & 0.12 & 0.52 & 0.13 & 0.92 & 0.79 & 0.13 & 0.11 & 0.09 \\
    \hline
    %-----------------------------------------------
    \end{tabular}
    \label{CC-results}
 % \vspace{-2mm}   
\end{table*}

\vspace{+0.05in}
\noindent {\bf Parameter Setting.}  We set the ratio of malicious clients $f/n$ to be $\{$2\%, 5\%, 10\%, 20\%$\}$ for FLAME, CC, and CC-B, and $\{$5\%, 10\%, 20\%, 30\%$\}$ for MDAM and {FLDetector}, consider their different defense performance. 
In each malicious client, we set the data poisoning rate as $50\%$, meaning 50\% of the client data are poisoned.  
In MDAM, we consider momentum coefficients $\beta = \{$0, 0.6, 0.9, 0.99$\}$, where $\beta=0$ means we do not use the momentum and it reduces to the standard MDA. In CC and CC-B, both the defense CC-$\tau$ and attack ATK-$\tau$ are set within $\{$0.1,1,10,100,1000$\}$. 
We also investigate the effect of Bucketing and set the  number of buckets $s$ in $\{$0,2,5,10$\}$, where $s=0$ means we do not use buckets. 

\vspace{+0.04in}
\noindent {\bf Evaluation Metric.} 
{For targeted backdoor poisoning attacks, we use both main task accuracy (MA) and backdoor accuracy (BA) as the evaluation metrics. 
An attack obtaining a \emph{larger} MA and a larger BA indicates it is more effective. 
For untargeted poisoning attacks, the goal is to reduce the main task accuracy. Therefore, a \emph{smaller} MA generally indicates better attack effectiveness.

\vspace{-2mm}
\subsection{Our Attack Results on FLAME, MDAM, {and FLDetector}}

\begin{table*}[!t]
%\footnotesize
\renewcommand{\arraystretch}{0.9}
\addtolength{\tabcolsep}{-1.2pt}
\caption{Results of attacking CC on IID FMNIST and CIFAR10, and attacking CC-B on non-IID FEMNIST with CC-$\tau$ = $0.1$, $1$, $100$, and $1000$. Our attack significantly outperforms LE and Fang when $f/n$ or $\tau$ is large.  
}
\centering
    \begin{tabular}{|c|c|c|c|c|c|c|c|c|c|c|c|c|c|c|c|c|c|c|}
    % \footnotesize
    \hline
    & & & & & \multicolumn{7}{c|}{Gradients known} &\multicolumn{7}{c|}{Gradients unknown} \\
    \cline{6-19}
    %\hline
    CC-$\tau$ & Dataset  & No attack & $f/n$ & LIE & \multicolumn{2}{c|}{AGR-tailored} & \multicolumn{5}{c|}{AGR-agnostic (ATK-$\tau$)} & \multicolumn{2}{c|}{AGR-tailored} & \multicolumn{5}{c|}{AGR-agnostic (ATK-$\tau$)} \\ % 
    \cline{6-19} %\cline{15-19}
    & &(MA) & ($\%$) &  & Fang & Ours & 0.1 & 1 & 10 & 100 & 1000 & Fang & Ours & 0.1 & 1 & 10 & 100 & 1000 \\
    \hline
    & & & 2 & 0.80 & 0.80 & 0.80 & 0.80 & 0.81 & 0.79 & 0.79 & 0.80 & 0.81 & 0.81 & 0.81 & 0.81 & 0.80 & 0.80 & 0.80 \\
    \cline{4-19}
    & FMNIST & 0.81 & 5 & 0.78 & 0.80 & 0.80 & 0.80 & 0.79 & 0.80 & 0.80 & 0.78 & 0.82 & 0.82 & 0.82 & 0.80 & 0.80 & 0.80 & 0.79 \\
    \cline{4-19}
    & & & 10 & 0.80 & 0.80 & 0.80 & 0.80 & 0.78 & 0.79 & 0.78 & 0.78 & 0.82 & 0.81 & 0.81 & 0.79 & 0.79 & 0.80 & 0.78 \\
    \cline{4-19}
    & &  & 20 & 0.81 & 0.80 & 0.79 & 0.79 & 0.77 & 0.78 & 0.76 & 0.77 & 0.80 & 0.79 & 0.79 & 0.76 & 0.75 & 0.73 & 0.75 \\
    \cline{2-19}
    & & & 2 & 0.12 & 0.10 & 0.11 & 0.11 & 0.10 & 0.12 & 0.13 & 0.09 & 0.12 & 0.10 & 0.10 & 0.14 & 0.15 & 0.12 & 0.12 \\
    \cline{4-19}
    0.1 & CIFAR10 & 0.12 & 5 & 0.12 & 0.13 & 0.08 & 0.08 & 0.10 & 0.09 & 0.10 & 0.13 & 0.11 & 0.10 & 0.10 & 0.12 & 0.09 & 0.12 & 0.09\\
    \cline{4-19}
    & &  & 10 & 0.10 & 0.10 & 0.10 & 0.10 & 0.11 & 0.09 & 0.10 & 0.08 & 0.11 & 0.10 & 0.10 & 0.12 & 0.12 & 0.09 & 0.11 \\
    \cline{4-19}
    & &  & 20 & 0.14 & 0.11 & 0.10 & 0.10 & 0.11 & 0.10 & 0.09 & 0.10 & 0.11 & 0.10 & 0.10 & 0.09 & 0.10 & 0.11 & 0.09 \\
    \cline{2-19}
    & & & 2 & 0.86 & 0.86 & 0.85 & 0.85 & 0.86 & 0.83 & 0.85 & 0.85 & 0.86 & 0.84 & 0.84 & 0.83 & 0.84 & 0.81 & 0.83 \\
    \cline{4-19}
    & EFMNIST & 0.87 & 5 & 0.85 & 0.85 & 0.84 & 0.84 & 0.83 & 0.85 & 0.85 & 0.81 & 0.85 & 0.83 & 0.83 & 0.84 & 0.82 & 0.80 & 0.81\\
    \cline{4-19}
    & & & 10 & 0.81 & 0.85 & 0.84 & 0.84 & 0.81 & 0.72 & 0.79 & 0.78 & 0.85 & 0.85 & 0.85 & 0.81 & 0.81 & 0.78 & 0.78 \\
    \cline{4-19}
    & &  & 20 & 0.83 & 0.84 & 0.86 & 0.86 & 0.77 & 0.72 & 0.67 & 0.69 & 0.85 & 0.82 & 0.82 & 0.76 & 0.64 & 0.60 & 0.64\\
    \hline
    %-----------------------------------------------
    & & & 2 & 0.85 & 0.84 & 0.83 & 0.84 & 0.83 & 0.83 & 0.82 & 0.83 & 0.85 & 0.84 & 0.84 & 0.84 & 0.83 & 0.83 & 0.82 \\
    \cline{4-19}
    & FMNIST & 0.84 & 5 & 0.84 & 0.84 & 0.82 & 0.84 & 0.82 & 0.81 & 0.81 & 0.82 & 0.85 & 0.80 & 0.84 & 0.80 & 0.81 & 0.80 & 0.80 \\
    \cline{4-19}
    & & & 10 & 0.84 & 0.83 & 0.79 & 0.84 & 0.79 & 0.79 & 0.78 & 0.79 & 0.84 & 0.80 & 0.84 & 0.80 & 0.78 & 0.78 & 0.77 \\
    \cline{4-19}
    & &  & 20 & 0.79 & 0.75 & 0.73 & 0.83 & 0.73 & 0.75 & 0.74 & 0.75 & 0.79 & 0.74 & 0.82 & 0.74 & 0.73 & 0.74 & 0.74\\
    \cline{2-19}
    & & & 2 & 0.38 & 0.36 & 0.33 & 0.39 & 0.33 & 0.37 & 0.34 & 0.31 & 0.40  & 0.38 & 0.34 & 0.38 & 0.34 & 0.31 & 0.32 \\
    \cline{4-19}
    1 & CIFAR10 & 0.40 & 5 & 0.37 & 0.39 & 0.30 & 0.39 & 0.30 & 0.32 & 0.21 & 0.24 & 0.40 & 0.40 & 0.38 & 0.40 & 0.32 & 0.30 & 0.28\\
    \cline{4-19}
    & &  & 10 & 0.40 & 0.34 & 0.38 & 0.38 & 0.38 & 0.18 & 0.13 & 0.12 & 0.36 & 0.34 & 0.34 & 0.34 & 0.23 & 0.12 & 0.07 \\
    \cline{4-19}
    & &  & 20 & 0.36 & 0.33 & 0.32 & 0.29 & 0.32 & 0.12 & 0.11 & 0.09 & 0.36 & 0.27 & 0.31 & 0.27 & 0.10 & 0.10 & 0.12\\
    \cline{2-19}
    & & & 2 & 0.92 & 0.93 & 0.92 & 0.93 & 0.92 & 0.89 & 0.90 & 0.91 & 0.92 & 0.91 & 0.92 & 0.91 & 0.90 & 0.90 & 0.89 \\
    \cline{4-19}
    & EFMNIST & 0.94 & 5 & 0.92 & 0.93 & 0.91 & 0.93 & 0.91 & 0.88 & 0.90 & 0.89 & 0.92 & 0.90 & 0.92 & 0.90 & 0.83 & 0.85 & 0.85 \\
    \cline{4-19}
    & & & 10 & 0.88 & 0.92 & 0.87 & 0.92 & 0.87 & 0.72 & 0.81 & 0.44 & 0.90 & 0.89 & 0.92 & 0.89 & 0.73 & 0.62 & 0.51 \\
    \cline{4-19}
    & &  & 20 & 0.85 & 0.87 & 0.85 & 0.92 & 0.85 & 0.11 & 0.10 & 0.11 & 0.90 & 0.82 & 0.91 & 0.82 & 0.09 & 0.09 & 0.11\\
    \hline
    %-----------------------------------------------
    & & & 2 & 0.84 & 0.72 & 0.49 &0.84 & 0.83 & 0.76 & 0.49 & 0.10 & 0.74 &  0.01 & 0.84 & 0.77 & 0.75 & 0.01 & 0.14\\
    \cline{4-19}
    & FMNIST & 0.85 & 5 & 0.84 & 0.72 & 0.13 &0.84 & 0.80 & 0.73 & 0.13 & 0.11 & 0.73 & 0.09 & 0.84 & 0.78 & 0.67 & 0.09 & 0.15 \\
    \cline{4-19}
    & & & 10 & 0.85 & 0.65 & 0.00& 0.83 & 0.78 & 0.62 & 0.00 & 0.00 & 0.69 & 0.00 & 0.84 & 0.79 & 0.60 & 0.00 & 0.00 \\
    \cline{4-19}
    & &  & 20 & 0.84 & 0.38 & 0.00&0.83 & 0.75 & 0.08 & 0.00 & 0.00 & 0.47 & 0.09 & 0.83 & 0.74 & 0.15 & 0.09 & 0.00\\
    \cline{2-19}
    & & & 2 & 0.63 & 0.66 & 0.09 & 0.63 & 0.64 & 0.63 & 0.09 & 0.13 & 0.62 & 0.10 & 0.62 & 0.63 & 0.63 & 0.10 & 0.12 \\
    \cline{4-19}
    100 & CIFAR10 & 0.64 & 5 & 0.64 & 0.69 & 0.10 & 0.64 & 0.34 & 0.23 & 0.10 & 0.11 & 0.61 & 0.15 & 0.64 & 0.60 & 0.44 & 0.15 & 0.12\\
    \cline{4-19}
    & & & 10 & 0.66 & 0.31 & 0.10 &  0.62 & 0.59 & 0.25 & 0.10 & 0.07 & 0.35 & 0.11 & 0.62 & 0.56 & 0.12 & 0.11 & 0.10 \\
    \cline{4-19}
    & &  & 20 & 0.63 & 0.20 & 0.10 & 0.60 & 0.51 & 0.16 & 0.10 & 0.08 & 0.22 & 0.13 & 0.59 & 0.57 & 0.15 & 0.13 & 0.10 \\
    \cline{2-19}     
    & & & 2 & 0.93 & 0.80 & 0.10 & 0.92 & 0.92 & 0.92 & 0.10 & 0.11 & 0.82 & 0.09 & 0.92 & 0.92 & 0.82 & 0.09 & 0.11 \\
    \cline{4-19}
    & EFMNIST & 0.92 & 5 & 0.88 & 0.79 & 0.11 &  0.91 & 0.92 & 0.10 & 0.11 & 0.11 & 0.78 & 0.11 & 0.92 & 0.92 & 0.11 & 0.11 & 0.09\\
    \cline{4-19}
    & & & 10 & 0.90 & 0.66 & 0.12 & 0.92 & 0.88 & 0.10 & 0.12 & 0.10 & 0.71 & 0.12 & 0.92 & 0.89 & 0.09 & 0.12 & 0.09 \\
    \cline{4-19}
    & &  & 20 & 0.84 & 0.42 & 0.05 & 0.91 & 0.86 & 0.11 & 0.05 & 0.07 & 0.53 & 0.09 & 0.92 & 0.11 & 0.10 & 0.09 & 0.10\\
    \hline    
    %-----------------------------------------------
    & & & 2 & 0.84 & 0.68 & 0.00 & 0.84 & 0.83 & 0.77 & 0.10 & 0.00 & 0.71 &  0.09 & 0.84 & 0.84 & 0.76 & 0.10 & 0.09 \\
    \cline{4-19}
    & FMNIST & 0.84 & 5 & 0.84 & 0.68 & 0.01 & 0.84 & 0.82 & 0.73 & 0.10 & 0.01 & 0.69 & 0.12 & 0.83 & 0.80 & 0.71 & 0.10 & 0.12  \\
    \cline{4-19}
    & & & 10 & 0.84 & 0.54 & 0.00 & 0.84 & 0.78 & 0.64 & 0.01 & 0.00 & 0.60 & 0.10 & 0.84 & 0.80 & 0.62 & 0.00 & 0.10  \\
    \cline{4-19}
    & &  & 20 & 0.84 & 0.23 & 0.10 &  0.83 & 0.11 & 0.09 & 0.00 & 0.10  & 0.31 & 0.10 & 0.83 & 0.76 & 0.10 & 0.00 & 0.10 \\
    \cline{2-19}
    & & & 2 & 0.63 & 0.51 & 0.10 & 0.63 & 0.64 & 0.57 & 0.17 & 0.10  & 0.55 & 0.09 & 0.64 & 0.64 & 0.60 & 0.11 & 0.09 \\
    \cline{4-19}
    1000 & CIFAR10 & 0.64 & 5 & 0.63 & 0.47 & 0.10 &  0.63 & 0.64 & 0.47 & 0.15 & 0.10  & 0.50 & 0.09 & 0.61 & 0.64 & 0.28 & 0.16 & 0.09\\
    \cline{4-19}
    & & & 10 & 0.63 & 0.45 & 0.01 & 0.63 & 0.64 & 0.13 & 0.10 & 0.01 & 0.48 & 0.11 & 0.62 & 0.64 &  0.14 & 0.13 & 0.11 \\
    \cline{4-19}
    & &  & 20 & 0.62 & 0.21 & 0.07 &  0.62 & 0.58 & 0.09 & 0.10 & 0.07 & 0.23 & 0.10 & 0.59 & 0.53 & 0.12 & 0.10 & 0.10\\
    \cline{2-19}
   
    & & & 2 & 0.85 & 0.69 & 0.09 &  0.92 & 0.91 & 0.84 & 0.10 & 0.09 & 0.74 & 0.12 & 0.92 & 0.92 & 0.88 & 0.09 & 0.12 \\
    \cline{4-19}
    & EFMNIST & 0.93 & 5 & 0.84 & 0.68 & 0.09 & 0.92 & 0.84 & 0.10 & 0.11 & 0.09  & 0.74 & 0.09 & 0.92 & 0.89 & 0.09 & 0.11 & 0.09\\
    \cline{4-19}
    & & & 10 & 0.85 & 0.49 & 0.10 &  0.92 & 0.87 & 0.11 & 0.12 & 0.10 & 0.50 & 0.09 & 0.92 & 0.88 & 0.12 & 0.12 & 0.09 \\
    \cline{4-19}
    & &  & 20 & 0.84 & 0.22 & 0.09 & 0.92 & 0.33 & 0.06 & 0.01 & 0.09& 0.23 & 0.09 & 0.91 & 0.85 & 0.10 & 0.09 & 0.09   \\
    \hline
    %-----------------------------------------------
    \end{tabular}
    \label{app:CC-results}
 \vspace{-1mm}   
\end{table*}

\begin{table*}[!t]
%\footnotesize
\renewcommand{\arraystretch}{0.9}
\addtolength{\tabcolsep}{1.1pt}
\caption{Impact of the number of buckets in CC-B on FEMNIST.}
\centering
    \begin{tabular}{|c|c|c|c|c|c|c|c|c|c|c|c|c|}
    % \footnotesize
    \hline
    $s$ & \multicolumn{3}{c|}{0} & \multicolumn{3}{c|}{2} & \multicolumn{3}{c|}{5} & \multicolumn{3}{c|}{10}\\ %  
    \hline
    ATK-$\tau$& No & 10 & 1000 & No & 10 & 1000 & No & 10 & 1000 & No & 10 & 1000 \\
    \hline
    CC-$\tau$ = $10$ & 0.92 & 0.12 & 0.11 & 0.92 & 0.09 & 0.11 & 0.92 & 0.15 & 0.03 & 0.92 & 0.10 & 0.06\\
    \hline
    CC-$\tau$ = $100$ & 0.91 & 0.10 & 0.09 & 0.92 & 0.09 & 0.08 & 0.92 & 0.11 & 0.09 & 0.91 & 0.12 & 0.11 \\
    \hline
    CC-$\tau$ = $1000$ & 0.92 & 0.11 & 0.09 & 0.93 & 0.10 & 0.10 & 0.92 & 0.10 & 0.09 & 0.91 & 0.12 & 0.10\\
    \hline
    \end{tabular}
    \label{bucketing-s}
    \vspace{-4mm}
\end{table*}
\noindent {\bf Comparing with a SOTA Targeted  Attack.} We choose the SOTA
Distributed Backdoor Attack (DBA) to FL \citep{xie2020dba}  as a baseline targeted poisoning attack for comparison. 
DBA decomposes a global trigger into several \emph{local} triggers and embeds these local triggers separately into the training data of different malicious clients. Compared with the classic \emph{centralized} backdoor  that injects the \emph{global} trigger, DBA is shown to be more persistent, stealthy, and effective (more details about DBA can be referred to \citep{xie2020dba}). 
To better show the attack effectiveness, our attack just uses the centralized backdoor, where we set the global trigger size to be the  same as that in DBA. Following DBA, we use the {rectangle pattern} as the global trigger and DBA separates the global trigger into four local triggers.  
We also show the DBA performance with different number of local triggers and the results are very close (see Table~\ref{app:DBA-results}). 
{The algorithm details of DBA and additional attack baselines are shown in Algorithms~\ref{DBA-algorithm}-\ref{Fang-algorithm}} in Appendix~\ref{app:SOTA}.

\begin{table*}[!t]
\centering
% \vspace{0mm}
\renewcommand{\arraystretch}{0.95}
\addtolength{\tabcolsep}{2pt}
\caption{Impact of dataset sizes on our attacks. Note: as the total number of clients increases, the dataset size per client decreases.}
%\vspace{-2mm}
\begin{tabular}{|c|c|c|c|c|c|c|c|c|c|}
\hline
Dataset & n  & FLAME & Ours & MDAM & Ours & FLDetector & Ours & CC(CC-B) & Ours \\ 
& & (MA) & (MA\ /\ BA)& (MA) & (MA\ /\ BA)& (MA) & (MA\ /\ BA)& (MA) & (MA) \\
\hline
  & 30 & 0.91  & 0.91\ /\ 0.95 & 0.93 & 0.93\ /\ 0.97 & 0.92       & 0.92\ /\ 0.98 & 0.82     & 0.41    \\
\cline{2-10}
     FMNIST   & 40 & 0.92  & 0.91\ /\ 0.95 & 0.92 & 0.92\ /\ 1.00 & 0.93       & 0.93\ /\ 0.95 & 0.83     & 0.37    \\
        \cline{2-10}
        & 50 & 0.92  & 0.92\ /\ 0.96 & 0.93 & 0.93\ /\ 0.99 & 0.93       & 0.93\ /\ 1.00 & 0.84     & 0.40    \\ 
\hline
 & 30 & 0.72  & 0.72\ /\ 0.78 & 0.72 & 0.71\ /\ 0.83 & 0.76       & 0.75\ /\ 0.87 & 0.68     & 0.10    \\ \cline{2-10}
      CIFAR10  & 40 & 0.70  & 0.72\ /\ 0.79 & 0.73 & 0.72\ /\ 0.84 & 0.74       & 0.73\ /\ 0.88 & 0.66     & 0.09    \\ \cline{2-10}
        & 50 & 0.71  & 0.72\ /\ 0.79 & 0.73 & 0.73\ /\ 0.81 & 0.75       & 0.75\ /\ 0.84 & 0.66     & 0.11    \\ 
\hline
 & 30 & 0.92  & 0.91\ /\ 0.92 & 0.95 & 0.95\ /\ 0.88 & 0.91       & 0.91\ /\ 0.75 & 0.91     & 0.12    \\ \cline{2-10}
     FEMNIST   & 40 & 0.92  & 0.92\ /\ 0.94 & 0.96 & 0.95\ /\ 0.91 & 0.90       & 0.90\ /\ 0.71 & 0.90     & 0.10    \\ \cline{2-10}
        & 50 & 0.94  & 0.91\ /\ 0.93 & 0.96 & 0.96\ /\ 0.89 & 0.90       & 0.90\ /\ 0.73 & 0.92     & 0.09    \\ 
\hline
\end{tabular}
\label{table:fl_analysis}
\end{table*}
%\vspace{-2mm}

\begin{figure*}[!t]
\centering	
\includegraphics[width=0.9\textwidth]{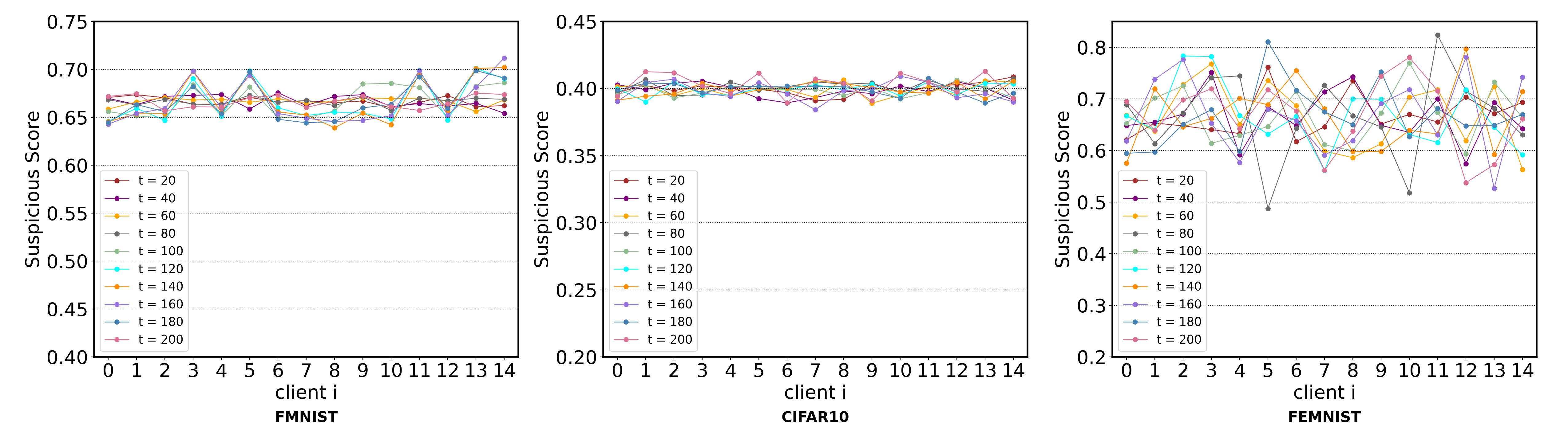}
     %\vspace{-2mm}
	\caption{{Suspicious scores per client and per FL round computed by FLDetector under our AGR-agnostic  and gradient-unknown attack. Here, clients $0-4$ are malicious and the remaining ones are benign, and $t$ is the FL training round.} We observe the suspicious scores are similar in all clients and FL rounds, hence making k-means clustering hard to detect the malicious scores.} 
        \label{fig:score-clients}
    %\vspace{-2mm}
\end{figure*}

\begin{figure*}[!t]
\centering	
\includegraphics[width=0.9\textwidth]{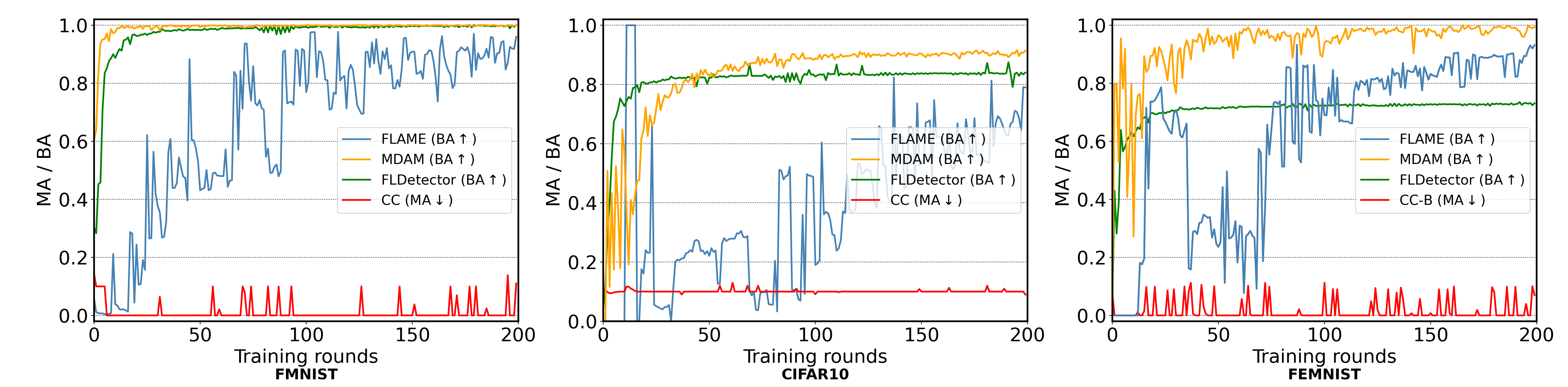}
     \vspace{-1mm}
	\caption{The impact of different numbers of training
rounds on our attacks. 
} 
        \label{fig:epoch-rebuttal}
    \vspace{-3mm}
\end{figure*}

\noindent {\bf Results on Attacking FLAME.}
The experimental results on different threat models are shown in Table \ref{FLAME-results}. we have the following key observations: 1) \emph{DBA is ineffective against FLAME, while our attack is effective.}  DBA obtains BAs that are  low. This shows FLAME can defend against DBA, which is also verified in \citep{nguyenflame}. In contrast, our attack achieves very high BAs, validating that it can break FLAME and our optimization-based attack is promising. 
2) \emph{Our attack can better maintain the FL performance than DBA.} Our attack obtains identical or better MAs than DBA and has closer MAs to those under no  attack. This implies our attack does not affect the main task performance.  
{3) \emph{In general, our attack has better performance under a strong threat model than that under a weak threat model.}} Specifically, comparing AGR-tailored vs. AGR-agnostic and gradients-known vs. gradients-unknown, all the BAs obtained by our attack are larger, while with similar MAs. 
4) \emph{Defending against attacks on non-IID datasets is more challenging than on IID datasets.} 
We notice that the BAs on the non-IID FEMNIST are much larger than those on the IID FMNIST and CIFAR10. This is because 
the client models can be more
diverse when trained on non-IID data,  thus making it
more difficult to differentiate between
benign models and malicious models via similarity metrics. 
5) \emph{More malicious clients yield better attack performance.} 
Our attack can obtain larger BAs, with an increasing number of malicious clients. This is obvious, since more malicious clients have a larger space to perform the attack. We notice that $5\%$-$10\%$ colluding malicious clients are sufficient to obtain a promising attack performance.  

\vspace{+0.1in}
\noindent {\bf Results on Attacking MDAM.}
The detailed results are in Table~\ref{MDAM-results} where we set the momentum coefficient to be $\beta=0.9$. The results on the impact of 
 $\beta$ 
 are shown in Table~\ref{app:MDAM-results}. 
We have similar observations as those on attacking FLMAE. For instance, our attack achieves larger MAs and BAs than DBA, showing our attack is more effective than DBA. 
MDAM can completely defend against DBA on IID datasets when the fraction of malicious clients is small. Also, defending against attacks on non-IID dataset is more challenging. Similarly, all these results show that our optimization-based attack can evade the filtering strategy in MDAM.

%----------------------------------------------------
\vspace{+0.05in}
\noindent {{\bf Results on Attacking FLDetector.} 
Table~\ref{FLDetector-results} shows the results of our  (AGR-agnostic and gradient-unknown) attack on FLDetector. 
The results reveal FLDetector almost fails to defend against our attack under the least adversary knowledge. 
To understand the underlying reason, we show in Figure~\ref{fig:score-clients} 
 the computed suspicious scores by FLDetector for each client and in each FL round. We observe the suspicious scores
are similar across all (malicious and benign) clients and FL rounds. Hence, it is challenging to use $k$-means to detect malicious clients based on these suspicious scores. 
}
%----------------------------------------------------

%\vspace{-2mm}
\subsection{Our Attack Results on CC and CC-B}

\noindent {\bf Comparing with SOTA Untargeted Attacks.} {We choose two SOTA untargeted poisoning attack methods, namely LIE \citep{baruch2019little} (AGR-agnostic) and Fang \citep{fang2020local} (AGR-tailored), for comparison.}  
Roughly speaking, LIE computes the average $\mu$ and standard deviation $\delta$ of the benign gradients, and computes a coefficient $z$ based on the total number of benign and malicious clients, and finally computes the malicious update as $\mu+z\delta$.
{Fang calculates the average $\mu$ of benign gradients and introduces a perturbation $\mathbf{g}^p = -$sign$(\mu)$. Ultimately, it computes a malicious update as $\mathbf{g}^c = \mu + \gamma\mathbf{g}^p$. The attack initiates a relatively large $\gamma$ 
and iteratively halves it until $\gamma$ yields promising attack performance.} 
The algorithm details of  LIE and Fang are in Algorithms \ref{LIE-algorithm}-\ref{Fang-algorithm} in Appendix~\ref{app:SOTA}. 
Note that, different from targeted poisoning attacks, our goal in this attack setting is to achieve as small MAs as possible. 

\vspace{+0.05in}
\noindent {\bf Results on Attacking CC.} 
As stated in Section~\ref{sec:untargetedattack}, CC has the only important parameter $\tau$. An AGR-tailored attack implies ATK-$\tau$ set by the adversary equals to the true CC-$\tau$, while an AGR-agnostic attack means ATK-$\tau$ and CC-$\tau$ are different. Table~\ref{CC-results} shows our attack  results on IID FMNIST and CIFAR10 where we set CC-$\tau$=10 and try different ATK-$\tau$'s. The attack results on other CC-$\tau$'s are shown in Table~\ref{app:CC-results}. 
We observe that: 1) {Our attack, even in the gradients-unknown setting}, is more effective than LIE and Fang with known gradients; 
2) Our attack obtains small MAs when the adversary knows the true CC-$\tau$; 3) By setting ATK-$\tau$ to be a larger value, e.g., 1000, our attack is always effective. This is because a larger ATK-$\tau$ can always make the adversary easier to satisfy the adversary objective in Equation \ref{attack-CC}. Such results guide the adversary to set a relatively large $\tau$ in practice. 

\vspace{+0.05in}
\noindent {\bf Results on Attacking CC with Bucketing.} Since the bucketing strategy is mainly to address the heterogeneous data issue across clients, we only evaluate our attack against CC-B on the non-IID FEMNIST. As shown in Table \ref{CC-results}, our attack can significantly reduce the MAs especially when ATK-$\tau$ is larger. 
This suggests our attack on non-IID datasets can also evade the aggregation of CC-B. 
We also explore the impact of the number of buckets $s$ on our attack. As shown in Table \ref{bucketing-s}, the results show that our attack is insensitive to $s$. 

%\vspace{-2mm}

\begin{table}[!t]
\renewcommand{\arraystretch}{0.9}
\addtolength{\tabcolsep}{-3pt}
% \vspace{+1mm}
\caption{Comparison results between our attacks utilizing Algorithm \ref{alg:attack} and directly solving $\gamma$ against FLAME, MDAM, FLDetector, and CC(CC-B).}
% \vspace{-3mm}
\centering
    \begin{tabular}{|c|c|c|c|c|}
    \hline
    { Dataset} (MA\ /\ BA) & FLAME & MDAM & FLDetector & CC(CC-B)\\
    \hline
    FMNIST(Direct) & 0.92\ /\ 0.96 & 0.93\ /\ 0.98 & 0.93\ /\ 0.99 & 0.42 \ /\ -\\ 
    \hline
    FMNIST(Ours) & 0.92\ /\ 0.96 & 0.93\ /\ 0.99 & 0.93\ /\ 1.00 & 0.40 \ /\ - \\
    \hline
    CITAF10(Direct) & 0.72\ /\ 0.79 & 0.73\ /\ 0.83 & 0.74\ /\ 0.85 & 0.09 \ /\ - \\
    \hline
    CIFAR10(Ours) & 0.72\ /\ 0.79 & 0.72\ /\ 0.81 & 0.75\ /\ 0.84 & 0.11 \ /\ -\\
    \hline
    FEMNIST(Direct) & 0.91\ /\ 0.94 & 0.96\ /\ 0.90 & 0.90\ /\ 0.74 & 0.12 \ /\ - \\
    \hline
    FEMNIST(Ours) & 0.91\ /\ 0.93 & 0.96\ /\ 0.89 & 0.90\ /\ 0.73 & 0.09 \ /\ -\\
    \hline
    \end{tabular}
    \label{table: directly solving}
%\vspace{-2mm}
\end{table}

\begin{figure*}[!t]
\centering	
\includegraphics[width=0.89\textwidth]{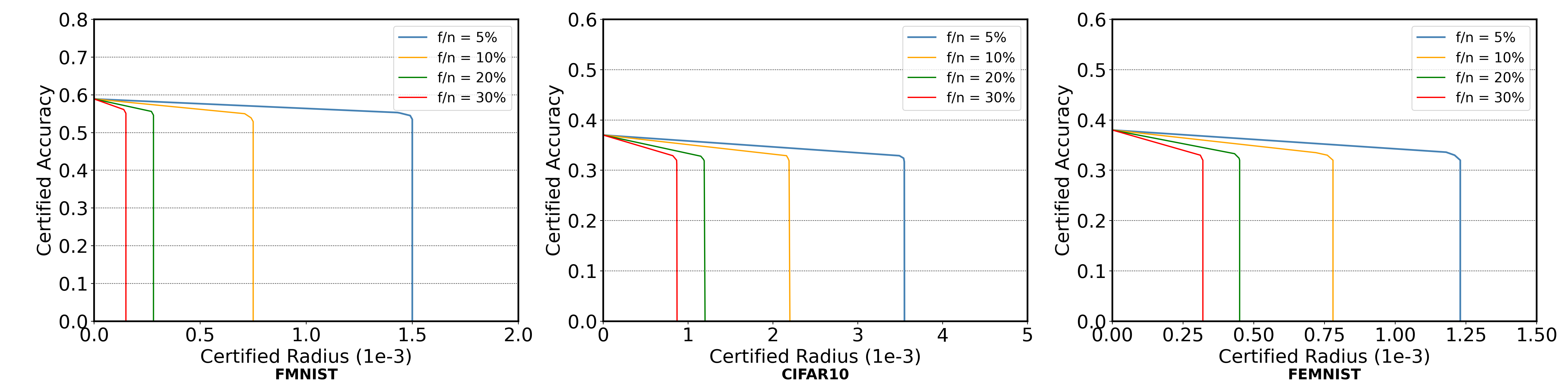}
    %\vspace{-2mm}
	\caption{{Certified accuracy of CRFL on our AGR-agnostic targeted poisoning attacks.}} 
        \label{fig:CRFL}
    % \vspace{-4mm}
\end{figure*}

\begin{figure*}[!t]
% \vspace{-3mm}
\centering	
\includegraphics[width=0.89\textwidth]{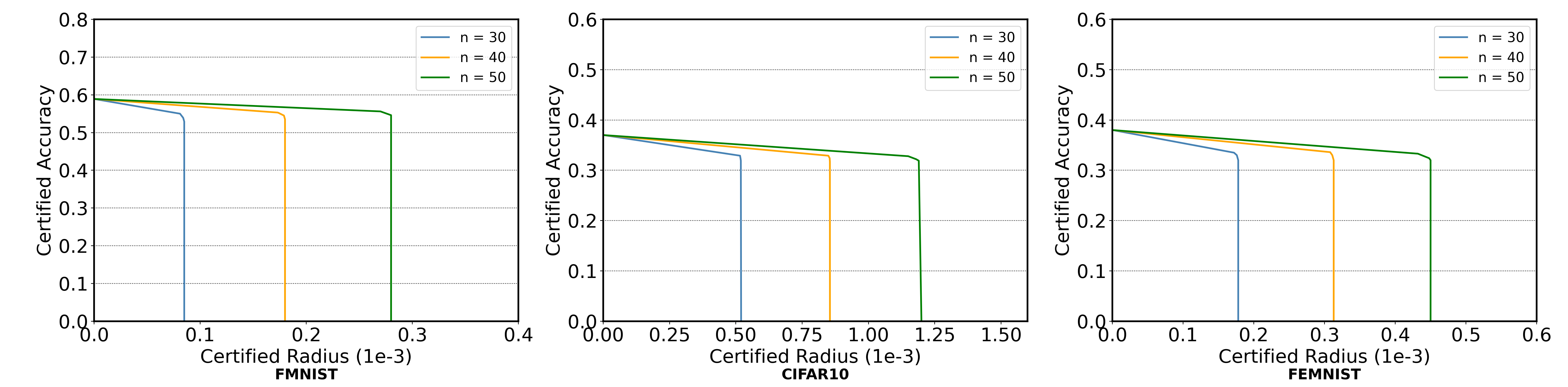}
    % \vspace{-4mm}
	\caption{{Certified accuracy of CRFL vs. the total number of clients $n$ on our AGR-agnostic targeted poisoning attacks.}} 
        \label{app:fig:CRFL-rebuttal}
\end{figure*}

\begin{figure*}[!t]
\centering	
% %\vspace{-2mm}
\includegraphics[width=0.89\textwidth]{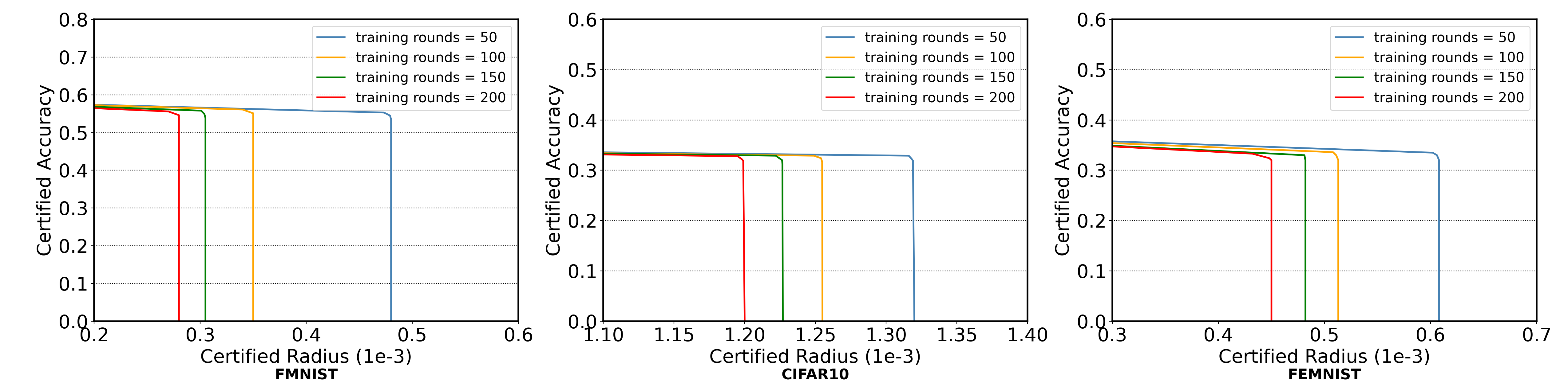}
    % \vspace{-4mm}
	\caption{{Certified accuracy of CRFL vs. the training rounds on our AGR-agnostic targeted poisoning attacks.}} 
        \label{app:fig:CRFL-epoch-rebuttal}
\end{figure*}

\subsection{More Experimental Results}
\noindent {\bf Impact of Training Rounds and Dataset Sizes.} We explore the impact of varying number of training rounds and dataset sizes on our attacks in the default setting. Figure \ref{fig:epoch-rebuttal} and Table \ref{table:fl_analysis} show the results on the two hyperparameters, respectively. We see the performance of most experiments (except FLAME) stabilizes after 100 rounds, and the proposed attacks are effective across different dataset sizes.

\vspace{+0.05in}
\noindent {\bf {Comparison of Directly Solving for $\gamma$.}} We also investigate the impact of the $\gamma$ solver. In these experiments, we evaluate our attacks by directly solving for $\gamma$. The comparison results under ``AGE-tailored + gradient known'' and the default setting are presented in Table \ref{table: directly solving}. We observe both two types of solvers produce very similar attack performance. On average, we find the difference of the calculated $\gamma$ between our iterative solver and the direct solver is less than 5\%, and the iterative solver is slightly faster (approximately $1.2 \times$) than the direct solver.

\vspace{+0.05in}
\noindent {\bf More attack baselines.} We test two more untarget attacks (sign flipping and label flipping attacks \citep{fang2020local}) against CC(-B), and a targeted attack \citep{wang2020attack} against MDAM. The results under the default setting are shown in Table \ref{table: defense 2}. We can see the SOTA robust FL can effectively defend against these attacks.
\begin{table}[!t]
\renewcommand{\arraystretch}{1.1}
\addtolength{\tabcolsep}{-6pt}
% \vspace{+2mm}
\footnotesize
\caption{Comparison results between ours and other 
baselines. }
%\vspace{-2mm}
\centering
    \begin{tabular}{|c|c|c|c|c|c|c|c|}
    \hline
    {Dataset}& CC(CC-B) & sign flip & label flip & Ours & MDAM & (Wang 2020) & Ours \\
    & (MA) & (MA) & (MA) & (MA) & (MA) & (MA\ /\ BA) & (MA\ /\ BA) \\
    \hline
    FMNIST & 0.84 & 0.69 & 0.72 & 0.40 & 0.93 & 0.93\ /\ 0.00 & 0.93\ /\ 0.45\\ 
    \hline
    CIFAR10 & 0.66 & 0.48 & 0.52 & 0.11 & 0.73 & 0.73\ /\ 0.01 & 0.72\ /\ 0.80 \\
    \hline
    FEMNIST & 0.92 & 0.73 & 0.87 & 0.09 & 0.96 & 0.96\ /\ 0.43 & 0.95\ /\ 0.78\\ 
    \hline
    \end{tabular}
    \label{table: defense 2}
    %\vspace{-2mm}
\end{table}

\vspace{+0.05in}
\noindent {\bf Comparison with class bias increase attack.} We evaluate the class bias increase attack against FLAME, MDAM, and FLDetector with a 50\%
bias level, where we randomly misclassify 50\% samples from not the target class (e.g., class 5) to the target class (e.g., class 7). The results
in the default setting are shown in Table \ref{app: table: Class Bias Attack}. We can see this attack fails to break SOTA AGRs, while our optimization-based attack
successfully bypasses these defenses.

\begin{table}[!t]
% \small 
\renewcommand{\arraystretch}{0.9}
%\addtolength{\tabcolsep}{1.1pt}
% \vspace{-5mm}
\caption{Comparison results between our attack and the class bias attack (CBA) against FLAME, MDAM, and FLDetector. }
% \vspace{-2mm}
\centering
\small 
    \begin{tabular}{|c|c|c|c|c|}
    \hline
    {\bf Dataset}& Attack & FLAME & MDAM & FLDetector \\
    \hline
    FMNIST & CBA & 0.92\ /\ 0.03 & 0.92\ /\ 0.01 & 0.93\ /\ 0.04\\ 
    \cline{2-5}
     & Ours  & 0.92\ /\ 0.96 & 0.93\ /\ 0.99& 0.93\ /\ 1.00 \\
    \hline
    CIFAR10 & CBA & 0.72\ /\ 0.02 & 0.72\ /\ 0.08 & 0.73\ /\ 0.02 \\ 
    \cline{2-5}
     & Ours  & 0.72\ /\ 0.79 &0.72\ /\ 0.81 & 0.75\ /\ 0.84 \\
    \hline
    FEMNIST & CBA & 0.92\ /\ 0.04 & 0.95\ /\ 0.01 & 0.90\ /\ 0.02 \\ 
    \cline{2-5}
     & Ours  & 0.91\ /\ 0.93 &0.96\ /\ 0.89 & 0.90\ /\ 0.73 \\
    \hline
    \end{tabular}
    \vspace{-2mm}
    \label{app: table: Class Bias Attack}
\end{table}

% \vspace{+0.5in}
\vspace{+1mm}
\noindent {\bf More defense baselines.} We mainly focus on robust aggregation based poisoning defenses, since they achieved SOTA defense performance. Here, we also test a popular non-robust aggregation based backdoor defense BaFFLe \citep{andreina2021baffle} against our gradient-unknown attack, and the results are shown in Table \ref{app: table: defense BaFFLe}. We can see BaFFLe is not effective enough, i.e., our attack still has high backdoor accuracy.

\begin{table}[!t]
\renewcommand{\arraystretch}{0.9}
\addtolength{\tabcolsep}{2pt}
\vspace{+3mm}
% \footnotesize
\caption{Our attacks vs. BaFFLe \citep{andreina2021baffle}.}
%\vspace{-2mm}
\centering
    \begin{tabular}{|c|c|c|c|}
    \hline
    {Dataset}& BaFFLe & AGR-tailored & AGR-agnostic \\
    & (MA) & (MA\ /\ BA) & (MA\ /\ BA) \\
    \hline
    FMNIST &0.93 &0.93\ /\ 0.90& 0.93\ /\ 0.78 \\
    \hline
    CIFAR10 & 0.73& 0.73\ /\ 0.79&0.73\ /\ 0.64\\
    \hline
    FEMNIST & 0.93 & 0.93\ /\ 0.91 & 0.93\ /\ 0.84 \\
    \hline
    \end{tabular}
    %\vspace{-2mm}
    \label{app: table: defense BaFFLe}
\end{table}

% \vspace{-2mm}
\noindent {\bf Comparison with more AGRs used \citep{shejwalkar2021manipulating}.} We compare our attack with the attack in \citep{shejwalkar2021manipulating} against the AGRs in \citep{shejwalkar2021manipulating} in the AGR-agnostic fashion. The results in the default setting are shown in Table \ref{app: table: [31]}. We can see both attacks have similar performance, though our attack is not specially designed for AGRs in \citep{shejwalkar2021manipulating}. 

\begin{table}[!t]
\renewcommand{\arraystretch}{0.9}
\addtolength{\tabcolsep}{-5pt}
%\vspace{-2mm}
% \vspace{2mm}
\footnotesize
\caption{Comparison of our attack with the attack in \citep{shejwalkar2021manipulating}. }
%\vspace{-2mm}
\centering
    \begin{tabular}{|c|c|c|c|c|c|c|c|c|}
    \hline
    Dataset & Attack &Krum &MKrum&Bulyan&TrMean&Median&AFA&FangT \\
    && (MA) & (MA) & (MA)  & (MA) & (MA) & (MA) & (MA)\\
    \hline
    FMNIST&(2021)&0.10&0.17&0.06&0.11&0.05&0.02&0.23\\
    \cline{2-9}
    & Ours	&0.15	&0.14&	0.10&	0.12&	0.09&	0.11	&0.27\\
    \hline
    CIFAR10&(2021)	&0.09&	0.31&	0.35&	0.33&	0.33&	0.22&	0.36\\
    \cline{2-9}
    & Ours	&0.12&	0.34&	0.33&	0.30&	0.31&	0.27&	0.32\\
    \hline
    FEMNIST&	(2021) &	0.01&	0.71&	0.20&	0.28&	0.22&	0.71&	0.80 \\
    \cline{2-9}
    & 	Ours&	0.07&	0.69	&0.29	&0.30&	0.22&	0.73&	0.82 \\ 
    \hline
    \end{tabular}
    \label{app: table: [31]}
\end{table}

% \vspace{2mm}

%\vspace{-2mm}
\subsection{Provable Defense Results against our Attack}
{In this experiment, we assess the effectiveness of the SOTA provable defense CRFL \citep{xie2021crfl} (Algorithm \ref{CRFL-algorithm} in Appendix~\ref{app:SOTA}) against our (AGR-agnostic) targeted poisoning attack. 
The results are depicted in Figure~\ref{fig:CRFL}. 
The certified radius means the maximal ($l_2$) norm of the trigger that can be injected to the clean data, such that the backdoored model still makes correct predictions; and the certified accuracy is the ratio of testing data under the given certified radius. We observe a notably small certified radius/accuracy in all datasets, and CRFL is unable to certify any testing data whose injected backdoor trigger is with an $l_2$ radius larger than $3.5e-3$. 
These results reveal inherent challenges of CRFL to 
defend against our attacks, even in the AGR-agnostic setting.} 

\vspace{+0.05in}
\noindent {\bf More results for our attacks against CRFL.} {We investigate the impact of the total number of clients $n$ and total training rounds on CRFL, while keeping other parameters the default values.} The results are in Figure \ref{app:fig:CRFL-rebuttal} and Figure \ref{app:fig:CRFL-epoch-rebuttal}, respectively. We can see the certified radius increases as $n$ increases (this reduces the attack effect) and training rounds decrease (attack is less persistent).

\begin{table}[!t]
%\footnotesize
% \vspace{+1mm}
\renewcommand{\arraystretch}{1.0}
\addtolength{\tabcolsep}{1.1pt}
\caption{DBA results on FLAME and MDAM ($\beta = 0.9$) with different number of local triggers. Here, we choose $f/n=10\%$ on FMNIST for simplicity.}
% \vspace{-3mm}
\centering
    \begin{tabular}{|c|c|c|c|}
    \hline
    \#local triggers & 2 & 3 & 4 \\
    \hline
    FLAME & 0.86\ /\ 0.02 & 0.86\ /\ 0.02 & 0.85\ /\ 0.01 \\ 
    \hline
    MDAM & 0.93\ /\ 0.01 & 0.93\ /\ 0.00 & 0.93\ /\ 0.01 \\ %
    \hline
    \end{tabular}
    \label{app:DBA-results}
 % \vspace{-2mm}   
\end{table}

\section{Related Work}
\noindent {\bf Poisoning Attacks to FL.} 
Poisoning attacks can be classified as targeted and untargeted attacks 
based on the adversary's goal. 
{Targeted poisoning attacks to FL~\citep{saha2020hidden,bagdasaryan2020backdoor,wang2020attack,xie2020dba,zhang2022neurotoxin} aim to misclassify the targeted inputs as the attacker desires, while maintaining the performance on clients' clean inputs.} For instance, backdoor attacks \citep{bagdasaryan2020backdoor,wang2020attack,xie2020dba} are a subset of targeted attacks, where an adversary (e.g., malicious clients) injects a backdoor into the target inputs and tampers with their labels as the adversary desired label during training. During testing, the trained backdoored global model misclassifies any test input with the backdoor as the desired attack label, while correctly classifying the clean testing inputs. 
In contrast, {untargeted poisoning attacks~\citep{baruch2019little,bhagoji2019analyzing,fang2020local,shejwalkar2021manipulating} aim to minimize the accuracy of the global model on test inputs, which can be implemented by data poisoning \citep{tolpegin2020data} or model poisoning {\citep{shejwalkar2021manipulating,fang2020local,bhagoji2019analyzing}.} 
For instance, \citep{shejwalkar2021manipulating} proposes an untargeted poisoning attack framework to mount optimal model poisoning attacks. Different from the existing attacks, we propose an attack framework that unifies both targeted and untargeted attacks to FL. }

\noindent {\bf Defenses against 
Poisoning Attacks to FL.} 
Existing (empirical) defenses focus on designing robust aggregators (AGRs) and they can be roughly classified as two categories: the ones that limit the attack effectiveness via clipping malicious gradients~\citep{karimireddy2021learning,nguyenflame,yin2018byzantine,bagdasaryan2020backdoor,nguyen2021flguard}; and the ones that weaken the contributions of malicious models by detecting and filtering them~\citep{farhadkhani2022byzantine,rieger2022deepsight,blanchard2017machine,xie2018generalized,munoz2019byzantine,andreina2021baffle}. Well-known AGRs include (Multi-) Krum~\citep{blanchard2017machine}, Bulyan~\citep{guerraoui2018hidden}, Trimmed-Mean (TM) ~\citep{xie2018generalized,yin2018byzantine}, Median~\citep{xie2018generalized,yin2018byzantine}, Minimum Diameter Averaging (MDA)~\citep{pillutla2019robust}, Adaptive Federated Averaging (AFA)~\citep{munoz2019byzantine}, and FLTrust~\citep{cao2020fltrust}. However, all these defenses are broken by the optimization-based adaptive ({untargeted poisoning}) attack proposed by \citep{shejwalkar2021manipulating}. 

To mitigate this attack, SOTA poisoning defenses incorporating novel robust AGRs have been proposed, where the 
representatives are     
CC \citep{karimireddy2021learning}, MDAM \citep{farhadkhani2022byzantine}, FLAME \citep{nguyenflame}, and {FLDetector \citep{zhang2022fldetector}}\footnote{FedRecover~\citep{cao2023fedrecover} bases on  FLDetector 
to first detect malicious clients  and then recover the global model.}.  
For instance, CC clips malicious gradients by a tunable threshold $\tau$ (hyper-parameter), while FLAME sets the clipping threshold through computing the median value of the Euclidean distance between the global model and local models. In contrast, MDAM chooses a subset of  $n-f$ clients (where $f$ is the total number of malicious clients) with the smallest diameter to filter out the malicious gradients. {Similar to MDAM, FLDetector utilizes historical information to predict gradient updates and identifies malicious clients by assessing discrepancies from the actual values.}

In the paper, we design an optimization-based attack framework to break all these SOTA poisoning defenses. 

\section{Conclusion}
In this paper, we study poisoning attacks to federated learning and aim to break state-of-the-art poisoning defenses that use \emph{robust} robust aggregators. 
To this end, we propose an optimization-based attack framework, under which we design customized attacks by uncovering the vulnerabilities of these robust aggregators. 
Our attacks are extensively evaluated on various threat models and benchmark datasets. The experimental results validate our attacks can break all these robust aggregators and deliver significantly stronger 
attack performance that the SOTA attacks. 
Potential future works include designing effective (provable) defenses, generalizing the proposed attack framework on federated learning to other application domains, e.g., graph data~\citep{wang2022graphfl}, and generalizing the attack to SOTA defenses for privacy~\citep{NoorbakhshZHW24,ArevaloNDHW24,LiHL0PH0Q22}.

\bibliographystyle{ACM-Reference-Format}
\small
\bibliography{cite}
\clearpage
\section{Detailed Implementations of SOTA Robust AGRs}
\label{app:code}
\begin{algorithm}[H]%h
\caption{AGR - FLAME} 
\label{FLAME-algorithm}
\begin{flushleft}
{\bf Input:} $n$, ${\bf w}^0$, $T$ $\rhd n$ is the number of clients, ${\bf w}^0$ is the initial global model parameters, and $T$ is the number of training iterations      
{\bf Output: ${\bf w}^T$} $\rhd {\bf w}^T$  is the global parameter after $T$ iterations
\end{flushleft}
\begin{algorithmic}[1]
\For{each training iteration $t$ in $[1,T]$} 
    \For{each client $i$ in $[1,n]$}
        \State  $\mathbf{g}_i^{t}  \gets $CLIENTUPDATE$( {\bf w}^{t-1}, D_i)$ $\rhd$ The aggregator sends ${\bf w}^{t-1}$ to Client $i$ who trains ${\bf w}^{t-1}$ using its data $D_i$ locally to achieve local gradient $\mathbf{g}_i^{t}$ and sends $\mathbf{g}_i^{t}$ back to the aggregator
    \EndFor
    \State $(c_{11}^{t},...,c_{nn}^{t}) \gets $cos$(\mathbf{g}_1^{t},...,\mathbf{g}_n^{t})$
    $\rhd$ $\forall i,j \in (1,...,n)$, $c_{ij}^{t}$ is the cosine distance between $\mathbf{g}_i^{t}$ and $\mathbf{g}_j^{t}$
    \State $(b_{1}^{t},...,b_{L}^{t})$ $\gets$ HDBSCAN$(c_{11}^{t},...,c_{nn}^{t})$ 
    $\rhd L$ is the number of admitted models, $b_l^{t}$ is the index of the $l$-th model
    \State $(e_{1}^{t},...,e_{n}^{t}) \gets \left \| ({\bf w}^{t-1},(\mathbf{g}_1^{t},...,\mathbf{g}_n^{t})) \right \|_2$ 
    $\rhd e_i^t$ is the Euclidean distance between ${\bf w}^{t-1}$ and $\mathbf{g}_i^{t}$
    $q^t \gets$ MEDIAN $(e_{1}^{t},...,e_{n}^{t})$ $\rhd q^t$ is the adaptive clipping bound at round $t$ 
    \For{each client $l$ in $[1,L]$} 
        \State   $\mathbf{g}_i^{t} \gets$ $\mathbf{g}_i^{t} \cdot$ min$(1, (q^t/e_{b_l}^{t}))$ $\rhd (q^t/e_{b_l}^{t})$ is the clipping parameter, and $\mathbf{g}_i^{t}$ is clipped by the adaptive clipping bound
    \EndFor
    \State ${\bf w}^{t} \gets {\bf w}^{t-1}-\eta{\textstyle \sum_{l=1}^{L}\mathbf{g}_i^{t}/L}+N(0,\sigma^2)$ $\rhd$ Sever aggregates parameters and adds noise, and then updates the global parameter as ${\bf w}^{t}$
\EndFor
\end{algorithmic}
\end{algorithm} 

\begin{algorithm}[H]%h
\caption{AGR - MDAM} 
\label{MDAM-algorithm}
\begin{flushleft}
{\bf Input:} $n$, ${\bf w}^0$, $\beta$, ${\bf m}^0$, $T$ $\rhd $ $n$ is the number of clients, ${\bf w}^0$ is the initial global model parameters, $\beta \in [0,1) $ is the momentum coefficient of all the clients, ${\bf m}^0=0$ is the initial momentum of each honest client, and $T$ is the number of training iterations

{\bf Output: ${\bf w}^T$} $\rhd {\bf w}^T$  is the global parameter after $T$ iterations
\end{flushleft}
\begin{algorithmic}[1]  
\For{each training iteration $t$ in $[1,T]$} 
    \For{each client $i$ in $[1,n]$} 
        \State $\mathbf{g}_i^{t} \gets $CLIENTUPDATE$( {\bf w}^{t-1}, D_i)$ $\rhd$ The aggregator sends ${\bf w}^{t-1}$ to Client $i$ who trains ${\bf w}^{t-1}$ using its data $D_i$ locally to achieve local gradient $\mathbf{g}_i^{t}$ 
        \State ${\bf m}^t_{i} \gets \beta{\bf m}^{t-1}_{i}+(1-\beta)\mathbf{g}_i^{t}$ $ \rhd$ Each honest client sends to the server the momentum ${\bf m}^t_{i}$
    \EndFor
    \State   $S^t \in \underset{ S \subset [n], |S|=n-f }{\arg\min} \left\{ \underset{i,j \in S}{\max}\left \|\mathbf{m}^t_i-\mathbf{m}^t_j  \right \|_2 \right\}$ $ \rhd$ Server first chooses a set $S^t$ of cardinality $n-f$ with the smallest diameter
    \State   ${\bf w}^{t} \gets {\bf w}^{t-1}- \frac{\eta}{n-f}\sum_{i \in S^t}{\bf m}^t_{i} $  $\rhd$ Server then updates the global parameter as ${\bf w}^{t}$
\EndFor
\end{algorithmic} 
\end{algorithm} 

\begin{algorithm}[H]
\caption{{AGR - FLDetector}} 
\label{FLDetector-algorithm}
\begin{flushleft}
{\bf Input:} $n$, ${\bf w}^0$, $N$, $T \rhd N$ is the number of past iterations, and $T$ is the number of training iterations

{\bf Output: ${\bf w}^T$} $\rhd {\bf w}^T$  is the global parameter after $T$ iterations
\end{flushleft}
\begin{algorithmic}[1] 
\For{each training iteration $t$ in $[1,T]$} 
    \For{each client $i$ in $[1,n]$} 
        \State   $\mathbf{g}_i^{t}  \gets $CLIENTUPDATE$( {\bf w}^{t-1}, D_i)$ 
        \State   $\hat{\mathbf{g}}_i^t \gets \mathbf{g}_i^{t-1}+\hat{H}^t({\bf w}_t-{\bf w}_{t-1})$
    \EndFor
    \State   $d^t \gets [\left \|\hat{\mathbf{g}}_1^t - \mathbf{g}_1^t \right \|_2,\left \|\hat{\mathbf{g}}_2^t - \mathbf{g}_2^t  \right \|_2,...,\left \|\hat{\mathbf{g}}_n^t - \mathbf{g}_n^t  \right \|_2]$
    \State   $s_i^t \gets \frac{1}{N} {\textstyle \sum_{r=0}^{N-1}} d_i^{t-r}/\left \|d^{t-r} \right \|_1$
    \State   Determine the number of clusters $k$ by Gap statistics.
    \If{$k>1$}
        \State   Perform $k$-means clustering based on the suspicious scores $s_i^t$ with $k \gets $ 2. $\rhd$ The clients in the cluster with smaller average suspicious score is benign.
    \EndIf
    \State   $\mathbf{g}^{t} \gets 0$
    \For{ each client $i$ in the benign cluster}
        \State   $\mathbf{g}^{t} \gets \mathbf{g}^{t} + \mathbf{g}_i^{t} $
    \EndFor
    \State   ${\bf w}^{t} \gets {\bf w}^{t-1}- \eta \mathbf{g}^{t}$ $ \rhd$ Server updates the global parameter as ${\bf w}^{t}$
\EndFor
\end{algorithmic}
\end{algorithm} 

\vspace{-3.5mm}
\begin{algorithm}[h]
\caption{AGR - CC} 
\label{CC-algorithm}
\begin{flushleft}
{\bf Input:} $n$, ${\bf w}^0$, $\tau$, $T \rhd 
\tau$ is a predefined clipping threshold 

{\bf Output: ${\bf w}^T$} $\rhd {\bf w}^T$  is the global parameter after $T$ iterations
\end{flushleft}
\begin{algorithmic}[1]
\For{ each training iteration $t$ in $[1,T]$ } 
    \For{ each client $i$ in $[1,n]$ } 
        \State   $\mathbf{g}_i^{t}  \gets $CLIENTUPDATE$( {\bf w}^{t-1}, D_i)$  
        \State   $\mathbf{g}_i^{t} \gets$ $\mathbf{g}_i^{t} \cdot$ min$(1, \frac{\tau}{\left \|\mathbf{g}_i^{t}  \right \|_2 } )$ $\rhd \tau$ is the clipping parameter
    \EndFor
    \State   ${\bf w}^{t} \gets {\bf w}^{t-1}- \eta\sum_{i \in [n]}\mathbf{g}_i^{t} $ 
\EndFor
\end{algorithmic}
\end{algorithm} 
\vspace{-3mm}

\begin{algorithm}[h]
\caption{AGR - CC-B} 
\label{CC-B-algorithm}
\begin{flushleft}
{\bf Input:} $n$, ${\bf w}^0$, $\tau$, $s$, $T \rhd 
\tau$ is a predefined clipping threshold, and $s$ is the number of buckets

{\bf Output: ${\bf w}^T$} $\rhd {\bf w}^T$  is the global parameter after $T$ iterations
\end{flushleft}
\begin{algorithmic}[1]     
\For{each training iteration $t$ in $[1,T]$} 
    \For{each client $i$ in $[1,n]$ } 
        \State   $\mathbf{g}_i^{t}  \gets $CLIENTUPDATE$( {\bf w}^{t-1}, D_i)$ 
    \EndFor
    \State   Pick random permutation $\pi$ of $[n]$
    \For{each parameter $i$ in $[ 1,\left \lceil  n/s\right \rceil ]$ } 
        \State    $\bar{\bf g}^t_i=\frac{1}{s} \sum_{k=(i-1)\cdot s+1}^{\min(n,i \cdot s)} \mathbf{g}^t_{\pi (k)}$ $\rhd$ Bucketing mixes the data from all clients
    \EndFor
    \State   $\bar{\bf g}^t_i \gets$ $\bar{\bf g}^t_i \cdot$ min$(1, \frac{\tau}{\left \|\bar{\bf g}^t_i  \right \|_2 } )$ $\rhd \tau$ is the clipping parameter     
    \State   ${\bf w}^{t} \gets {\bf w}^{t-1}- \eta\sum_{i \in [n]}\bar{\bf g}^t_i $ 
\EndFor
\end{algorithmic} 
\end{algorithm} 

\section{More Details about the DBA, LIE, and Fang Attacks and the CRFL defense}
\label{app:SOTA}
\begin{algorithm}[H]
\caption{A baseline targeted poisoning attack: DBA} 
\label{DBA-algorithm}
\begin{flushleft}
{\bf Input:} $n$, ${\bf w}^0$, $n_{tri}$, $f$, $T \rhd$ $n$ is the number of clients, ${\bf w}^0$ is the initial global model parameters, $n_{tri}$ is the number of local triggers, $f$ is the number of adversaries, and $T$ is the number of training iterations 
 
{\bf Output: ${\bf w}^T$} $\rhd {\bf w}^T$  is the global parameter after $T$ iterations
\end{flushleft}
\begin{algorithmic}[1]   
\State Decompose a global trigger into $n_{tri}$ local triggers
\For{each training iteration $t$ in $[1,T]$ } 
    \For{each benign client $i$ in $[f+1,n]$ } 
        \State   $\mathbf{g}_i^{t}  \gets $CLIENTUPDATE$( {\bf w}^{t-1}, D_i)$  $\rhd$ 
        The aggregator sends ${\bf w}^{t-1}$ to Client $i$ who trains ${\bf w}^{t-1}$ using its data $D_i$ locally to achieve local gradient $\mathbf{g}_i^{t}$ 
    \EndFor
    \For{each adversary $j$ in $[1,f]$}
        \State   $D'_j \gets$ POISONING $(D_j, j$ mod $n_{tri})$  $\rhd$ Adversary $j$ poisons his data $D_j$ with ($j$ mod $n_{tri}$)-th local trigger  
        \State   $\mathbf{g}_j^{t}  \gets $CLIENTUPDATE$( {\bf w}^{t-1}, D'_j)$ $\rhd$ Adversary $j$ trains using its poisoning data $D'_j$ to achieve malicious gradient $\mathbf{g}_j^{t}$ 
    \EndFor
    \State   ${\bf w}^{t} \gets {\bf w}^{t-1}- \eta\sum_{i \in [n]}\mathbf{g}_i^{t} $ $\rhd$ Server updates the global parameter as ${\bf w}^{t}$
\EndFor
\end{algorithmic}
\end{algorithm}

\vspace{-2mm}
\begin{algorithm}[h]
\caption{A baseline untargeted poisoning attack: LIE} 
\label{LIE-algorithm}
\begin{flushleft}
{\bf Input:} $n$, ${\bf w}^0$, $f$, $T \rhd$ $n$ is the number of clients, ${\bf w}^0$ is the initial global model parameters, $f$ is the number of adversaries, and $T$ is the number of training iterations
 
{\bf Output: ${\bf w}^T$} $\rhd {\bf w}^T$  is the global parameter after $T$ iterations
\end{flushleft}
\begin{algorithmic}[1]   
\For{each training iteration $t$ in $[1,T]$} 
    \For{each benign client $i$ in $[f+1,n]$ } 
        \State   $\mathbf{g}_i^{t}  \gets $CLIENTUPDATE$( {\bf w}^{t-1}, D_i)$ $\rhd$  
        The aggregator sends ${\bf w}^{t-1}$ to Client $i$ who trains ${\bf w}^{t-1}$ using its data $D_i$ locally to achieve local gradient $\mathbf{g}_i^{t}$
    \EndFor
    \For{ each adversary $j$ in $[1,f]$ }
        \State   $s \gets \left \lfloor \frac{n}{2}+1  \right \rfloor -f$ 
        \State   $z \gets \max_z (\phi(z) < \frac{n-f-s}{n-f} )$ $ \rhd$ Adversary $j$ computes a coefficient $z$ based on the total number of benign and malicious clients, where $\phi(z)$ is the cumulative standard normal function 
        \State   $\mu \gets$ mean $(\mathbf{g}_{f+1}^{t},..., \mathbf{g}_n^{t}) $ $\rhd$ Compute the average $\mu$ of the benign gradients
        \State   $\sigma \gets$ std $(\mathbf{g}_{f+1}^{t},..., \mathbf{g}_n^{t}) $ $\rhd$ Compute the standard deviation $\sigma$ of the benign gradients 
        \State   $\mathbf{g}_{j}^{t} \gets \mu+z\sigma $ $\rhd$ Update the malicious gradient
    \EndFor
    \State   ${\bf w}^{t} \gets {\bf w}^{t-1}- \eta\sum_{i \in [n]}\mathbf{g}_i^{t} $  $\rhd$ Server updates the global parameter as ${\bf w}^{t}$
\EndFor
\end{algorithmic}
\end{algorithm} 

\begin{algorithm}[t]
\caption{{A baseline untargeted poisoning attack: Fang}} 
\label{Fang-algorithm}
\begin{flushleft}
{\bf Input:} $n$, ${\bf w}^0$, $\varepsilon $, $f$, $T \rhd$ $n$ is the number of clients, ${\bf w}^0$ is the initial global model parameters, $\varepsilon$ is the threshold of updating malicious gradients, $f$ is the number of adversaries, and $T$ is the number of training iterations 
 
{\bf Output: ${\bf w}^T$} $\rhd {\bf w}^T$  is the global parameter after $T$ iterations
\end{flushleft}
\begin{algorithmic}[1]
\For{each training iteration $t$ in $[1,T]$} 
    \For{each benign client $i$ in $[f+1,n]$} 
        \State   $\mathbf{g}_i^{t}  \gets $CLIENTUPDATE$( {\bf w}^{t-1}, D_i)$ $\rhd$ Client $i$ trains using its benign data $D_i$ locally  
    \EndFor
    \State   $\mu \gets$ mean $(\mathbf{g}_{f+1}^{t},..., \mathbf{g}_n^{t}) \rhd$ Compute the average $\mu$ of the benign gradients
    \For{each adversity $j$ in $[1,f]$ }
        \While{$\gamma >\varepsilon $}
            \State  $\mathbf{g}^p \gets -$sign$(\mu)$ 
            \State  $\mathbf{g}_j^t \gets \mu + \gamma\mathbf{g}^p$
            \State  $\gamma = \gamma/2 $
        \EndWhile
    \EndFor
    \State   ${\bf w}^{t} \gets {\bf w}^{t-1}- \eta\sum_{i \in [n]}\mathbf{g}_i^{t} $ $\rhd$ Server updates the global parameter as ${\bf w}^{t}$
\EndFor
\end{algorithmic}
\end{algorithm}

\begin{algorithm}[H]%h
\caption{Provable defense - CRFL} 
\label{CRFL-algorithm}
\begin{flushleft}
{\bf Input:} $x_{test}$, $y_{test}$, ${\bf w}^T$, $\rho_T$, $h(\cdot,\cdot) \rhd$ a test sample $x_{test}$ with true label $y_{test}$, the global model parameters ${\bf w}^T$, clipping threshold $\rho_T$, and the classifier $h$

{\bf Output:} $\hat{c}_A$, RAD $\rhd \hat{c}_A$ is the prediction and RAD is the certified radius
\end{flushleft}
\begin{algorithmic}[1]
\For{$k = 0, 1, \cdots, M$ } 
    \State  $\epsilon_T^k \gets$ a sample drawn from $\mathcal{N}(0,\sigma_T^2\mathbf{I})$ 
    \State  
    $\widetilde{\omega}_T^k = {\bf w}^T / max(1,\frac{{\bf w}^T}{\rho_T}) + \epsilon_T^k$
\EndFor
\State $\rhd$ Calculate empirical estimation of $p_A$, $p_B$ for $x_{test}$
\State  
$counts \gets$ GetCounts($x_{test},\{\widetilde{\omega}_T^1, \cdots,  \widetilde{\omega}_T^M\}$)
\State  $\hat{c}_A$, $\hat{c}_B \gets$ top two indices in $counts$%; \quad 
\State  
$\hat{p}_A$, $\hat{p}_B \gets counts[\hat{c}_A]/M, counts[\hat{c}_B]/M$
\State $\rhd$ Calculate lower and upper bounds of $ {p}_A$,${p}_B$
\State  
$\underline{p_A}$, $\overline{{p}_B} \gets$ CalculateBound$(\hat{p}_A,\hat{p}_B,N,\alpha)$
\If{$\underline{p_A} > \overline{{p}_B}$}
    \State  RAD $=$ CalculateRadius$(\underline{p_A}, \overline{{p}_B}) \rhd$ According to Corollary 1 in CRFL \citep{xie2021crfl}
\Else
    \State RAD $= 0$
    \State $\hat{c}_A=$ ABSTAIN
\EndIf
\end{algorithmic}
\end{algorithm} 
\end{document}